\documentclass[epj]{svjour}
\usepackage{graphics}
\usepackage{graphicx}
\usepackage{float}
\usepackage{bbold}
\usepackage{xcolor,rotating}
\usepackage{bbm}
\usepackage{dsfont}

\usepackage{booktabs}
\usepackage{multirow}
\usepackage{ctable}
\usepackage{xspace}
 \usepackage{subfigure}
 \usepackage{bm}
 \usepackage{siunitx}

\DeclareSIUnit\c{\textit{c}}
\DeclareSIUnit[per-mode=symbol]\mevc{\MeV\per\c}
\DeclareSIUnit[per-mode=symbol]\gevc{\GeV\per\c}
\DeclareSIUnit[per-mode=symbol]\evcc{\eV\per\c\squared}
\DeclareSIUnit[per-mode=symbol]\kevcc{\keV\per\c\squared}
\DeclareSIUnit[per-mode=symbol]\mevcc{\MeV\per\c\squared}
\DeclareSIUnit[per-mode=symbol]\gevcc{\GeV\per\c\squared}

\def\0#1#2{\frac{#1}{#2}}

\renewcommand{\figurename}{{\bf Figure}}
\renewcommand{\thefigure}{{\bf\arabic{figure}}}
\renewcommand{\tablename}{{\bf Table}}
\renewcommand{\thetable}{{\bf\arabic{table}}}

\newcommand\T{\rule{0pt}{2.6ex}}       
\newcommand\B{\rule[-1.2ex]{0pt}{0pt}} 

\newcommand{\mult}       {charged-particle multiplicity\xspace}

\newcommand{\mpi}        {multiparton interactions\xspace}

\newcommand{\Nmpi}        {\ensuremath{N_\mathrm{MPI}}\xspace}
\newcommand{\Nch}        {\ensuremath{N_\mathrm{ch}}\xspace}

\newcommand{\dndeta}        {\ensuremath{\D N_\mathrm{ch}/ \D \eta}\xspace}
\newcommand{\D}          {\ensuremath{\mathrm{d}}\xspace}

\newcommand{\jpsi}       {{\ensuremath{\mathrm{J}/\psi }}\xspace}

\newcommand{\pt}         {\ensuremath{p_\mathrm{T}}\xspace}

\newcommand{\fig}[1]     {Fig.~\ref{#1}}

\usepackage{hyperref}
\hypersetup{
colorlinks=true,
citecolor=blue,
citebordercolor=red,
linktoc=all,
linkcolor=blue
}

\def\0#1#2{\frac{#1}{#2}}

\renewcommand{\figurename}{{\bf Figure}}
\renewcommand{\thefigure}{{\bf\arabic{figure}}}
\renewcommand{\tablename}{{\bf Table}}
\renewcommand{\thetable}{{\bf\arabic{table}}}

\usepackage{amssymb}

\begin{document}

\title{Elucidating the multiplicity dependence of \jpsi production in proton-proton collisions with PYTHIA8}
\author{S. G. Weber\inst{1,2}, A. Dubla\inst{2,3}, A. Andronic\inst{4} \and A. Morsch\inst{5}}                     
\offprints{}          
\institute{Institut f{\"u}r Kernphysik, Technische Universit{\"a}t Darmstadt, 64289 Darmstadt, Germany \and GSI Helmholtzzentrum f{\"u}r Schwerionenforschung, 64248 Darmstadt, Germany \and Physikalisches Institut, Universit{\"a}t Heidelberg, 69120 Heidelberg, Germany \and Institut f{\"u}r Kernphysik, Westf\"{a}lische Wilhelms-Universit\"{a}t M\"{u}nster, 48149 M\"{u}nster, Germany, \and CERN, 1211 Geneva 23, Switzerland}
\date{Received: date / Revised version: date}
%


\abstract{
A study  of prompt and non-prompt \jpsi production as a function of charged-particle multiplicity in inelastic proton--proton (pp) collisions at a centre-of-mass energy of $\sqrt{s}$ = 13 TeV based on calculations using the PYTHIA8 Monte Carlo is reported.
Recent experimental data shows an intriguing stronger-than-linear increase of the self-normalized \jpsi yield with multiplicity; several models, based on initial or final state effects, have been able to describe the observed behaviour.
In this paper, the microscopic reasons for this behaviour, like the role of multiple parton interactions, colour reconnections and auto-correlations are investigated.
It is observed that the stronger-than-linear increase and the transverse momentum (\pt) dependence, contrary to what is predicted by the other available models, can be attributed to auto-correlation effects only. In absence of auto-correlation effects, the increase of the yield of \jpsi with multiplicity -- and in general for all hard processes -- is weaker than linear for multiplicities exceeding about three times the mean multiplicity. The possibility of disentangling auto-correlation effects from other physical phenomena by measuring the charged-particle multiplicity in different pseudo-rapidity and azimuthal regions relative to the \jpsi direction is investigated. In this regard, it is suggested to extend the experimental measurements of \jpsi production as a function of the charged-particle multiplicity by determining the multiplicity in several azimuthal regions and in particular in the Transverse region with respect to the direction of the \jpsi meson.
} 
     
\authorrunning{S. G. Weber et al.}
\titlerunning{Multiplicity dependence of \jpsi production in pp collisions with PYTHIA8}
\maketitle

\section{Introduction}
\label{intro}

Hadronic charmonium production at collider energies is a complex and not yet fully understood process. It
involves partonic interactions with large momentum transfer (hard processes), i.e. the initial heavy quark pair production, which can be described
by means of perturbative Quantum Chromodynamics (pQCD), as well as soft scale processes, i.e. the subsequent binding into a charmonium state. A comprehensive description of the transverse momentum and rapidity dependent production down to transverse momentum $\pt$=0 was recently achieved within the non-relativistic
QCD (NRQCD) formalism combined with a Colour Glass Condensate
(CGC) description of the incoming protons \cite{nrqcd_cgc}; polarization is also well described \cite{Ma:2018qvc}.
The correlation of charmonium (and also of heavy quarks in general) production with the charged particle multiplicity is of high interest, as it could give new insight into the interplay between hard and soft mechanisms in particle production, both at parton level and at hadronization.

In pp collisions at $\sqrt{s}$ = 7 TeV, ALICE has performed multiplicity dependent measurements  of inclusive
\jpsi production at mid and forward rapidity \cite{alice_jpsi_mult_7tev}, and prompt \jpsi, non-prompt \jpsi and D-mesons production at mid-rapidity \cite{alice_d_mult_7tev}.
The general observation is an increase of open and hidden charm production with 
multiplicity. For \jpsi production, multiplicities of about 4 times the mean value observed in minimum bias events are reached. The results are consistent with a linear, or stronger than linear increase. For D-meson production, 
multiplicities of about 6 times the mean value are reached, with a stronger than linear increase at the highest multiplicities.
Similar observations have been made by CMS for $\Upsilon$(nS) mesons at mid-rapidity: a linear increase
with the event activity measured at forward rapidity and a stronger-than-linear increase with the event activity measured at mid-rapidity was observed \cite{cms_y_mult}.

Different theoretical models attribute the observed behaviour to different underlying processes, such as the percolation mechanism \cite{percolation1}, higher Fock states in the proton  \cite{kopeliovich},  effects from the colour glass condensate EFT \cite{cgc_jpsi}, and multi-pomeron interaction combined with high-density effects (EPOS3 event generator \cite{epos3}).

In this article, we study prompt and non-prompt \jpsi production as a function of the charged-particle multiplicity in proton--proton (pp) collisions at a centre-of-mass energy of $\sqrt{s}$ = 13 TeV. The study is based on Monte Carlo (MC) event samples generated using PYTHIA8.
It is worth noting that the presented observations for non-prompt \jpsi production (\jpsi from heavy flavor hadron decays) are equally valid for open heavy-flavour hadrons in general, and the findings for prompt \jpsi apply also to bottomonium production.

\section{The PYTHIA event generator}
\label{pythiaintro}

PYTHIA \cite{pythia64,pythia8}  is an event generator for collisions of protons, leptons and nuclei. It has a complex physics model with a multitude of different processes implemented at different stages of the collision.
Proton-proton collisions contain one or more perturbative scattering processes between the incoming partons implemented within the MPI framework \cite{mpi}.
A typical event at LHC energies contains roughly between four and ten partonic interactions (PI)
\cite{beam_remnants}.
The number of PI per collision depends on the matter overlap in the collisions and, hence, on the pp impact parameter. 
The perturbative scattering processes are accompanied by Initial-State Radiation (ISR) and Final-State Radiation (FSR). 

Hadronization is implemented according to the Lund string fragmentation model \cite{lund_string}. The created partons and the beam-remnants are connected via colour fluxtubes, or strings storing  potential energy.
As the partons move apart the string breaks, producing light quark-antiquark pair. The process repeats itself, until small enough pieces of strings remain, which are then identified with on-shell hadrons. 
In the Colour Reconnection (CR) scenario \cite{color_reconnection}, strings can be rearranged between partons, so as to reduce the total string length. Partons from different PI can become connected to each other.
The reduction of the total string length leads to a reduction of the total multiplicity, since the bulk of particles are produced from the string breaking mechanism. 

The results reported in this paper are obtained from simulated non-diffractive events using PYTHIA version 8.230 \cite{pythia82} with the default Monash 2013 tune \cite{monash}.

\section{Charged-particle multiplicity}

\begin{figure}[th!]
  \centering
    \includegraphics[scale=0.4]{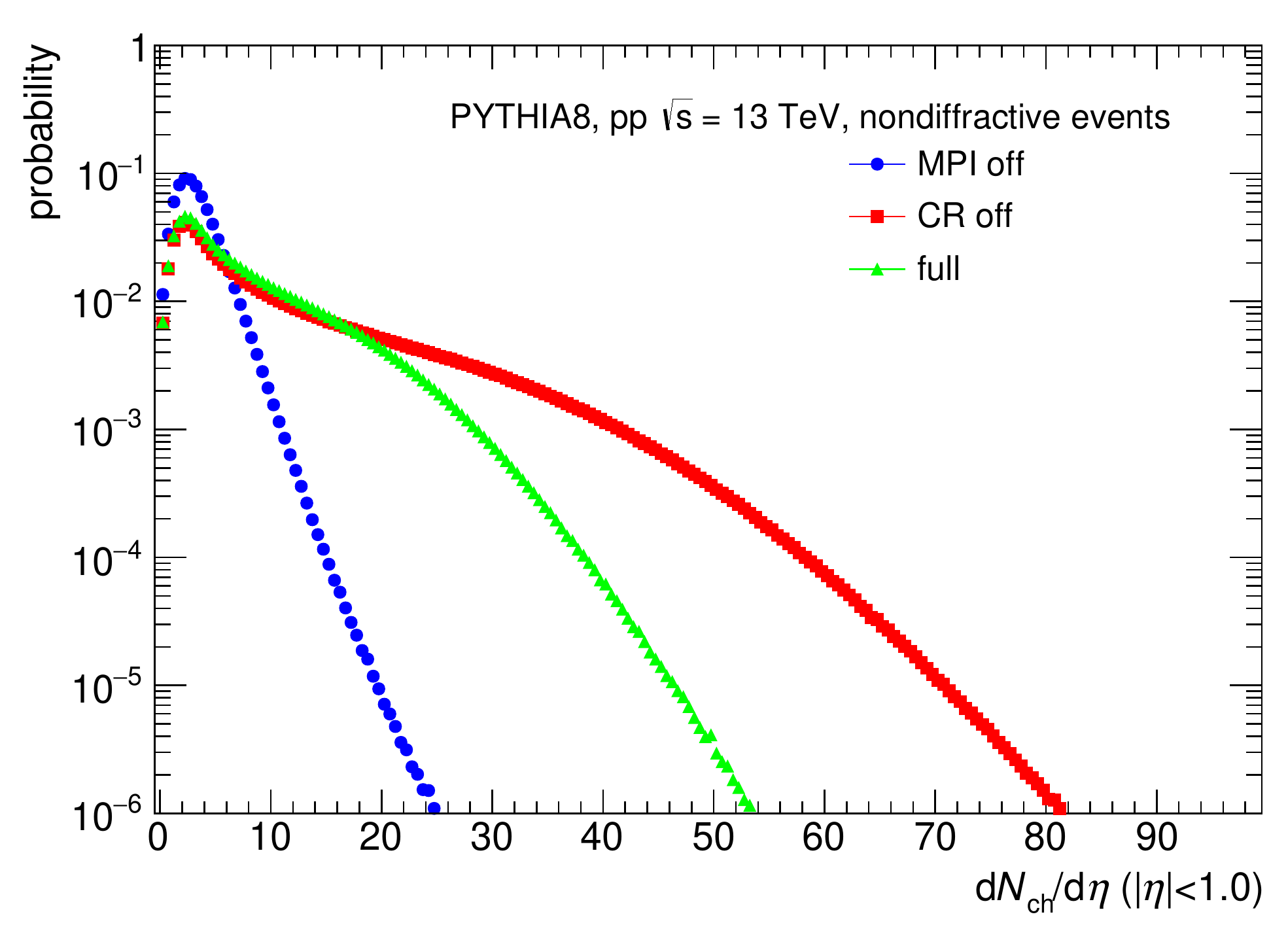} 
      \caption{ Charged-particle multiplicities at mid-rapidity $|\eta| < 1$ for non-diffractive pp collisions at $\sqrt{s}$ = 13 TeV in PYTHIA8 with default settings, CR off, and MPI off. }
          \label{fig_pythia_Nmpi}
  \end{figure}

The probability distributions of the charged particle at mid-rapidity ($|\eta| < 1$) for the default settings, for CR switched-off, and for MPI switched-off are shown in \fig{fig_pythia_Nmpi}. 
Multiplicity is defined as the number of primary charged-particles, according to the ALICE definition \cite{alice_primaries}. Note that the dependence of \jpsi production on the multiplicity at mid-rapidity  was investigated, as well as at forward rapidity, i.e. at the pseudo-rapidity of the V0 \cite{VZERO} detectors ($2.8 < \eta < 5.1$ and $-3.7 < \eta < -1.7$) of the ALICE apparatus.

The inclusion of MPI increases the average multiplicity by about a factor of 3 and the distribution becomes much wider. On the other hand the colour reconnection mechanism reduces the average multiplicity by about $30\%$ and also makes the distribution narrower. From earlier analysis it is known that the full simulation including both MPI and CR reproduces the \mult distribution measured at LHC reasonably well \cite{atlas_nch_13tev}.
Table \ref{tab1} lists the mean values and the RMS of the multiplicity distribution for inelastic non-diffractive collisions.

   \begin{table} [ht!]
   \centering
      \begin{tabular} {l|ccc}
          \hline
      & MPI off & CR off & full\B\\
    \hline
   $\langle\dndeta_{|\eta|<1.0}\rangle$    & 2.57 & 8.44  & 6.04 \T\\\
   RMS $\dndeta_{|\eta|<1.0}$            & 2.04 & 10.25 & 6.46 \B\\
    \hline
    \end{tabular}
  \caption{ Mean and RMS values of the charged-particle multiplicities distributions. }
    \label{tab1}
\end{table} 

\begin{figure*}[ht!]
\begin{minipage}[b]{0.5\linewidth}
\centering
    \includegraphics[width=0.9\linewidth]{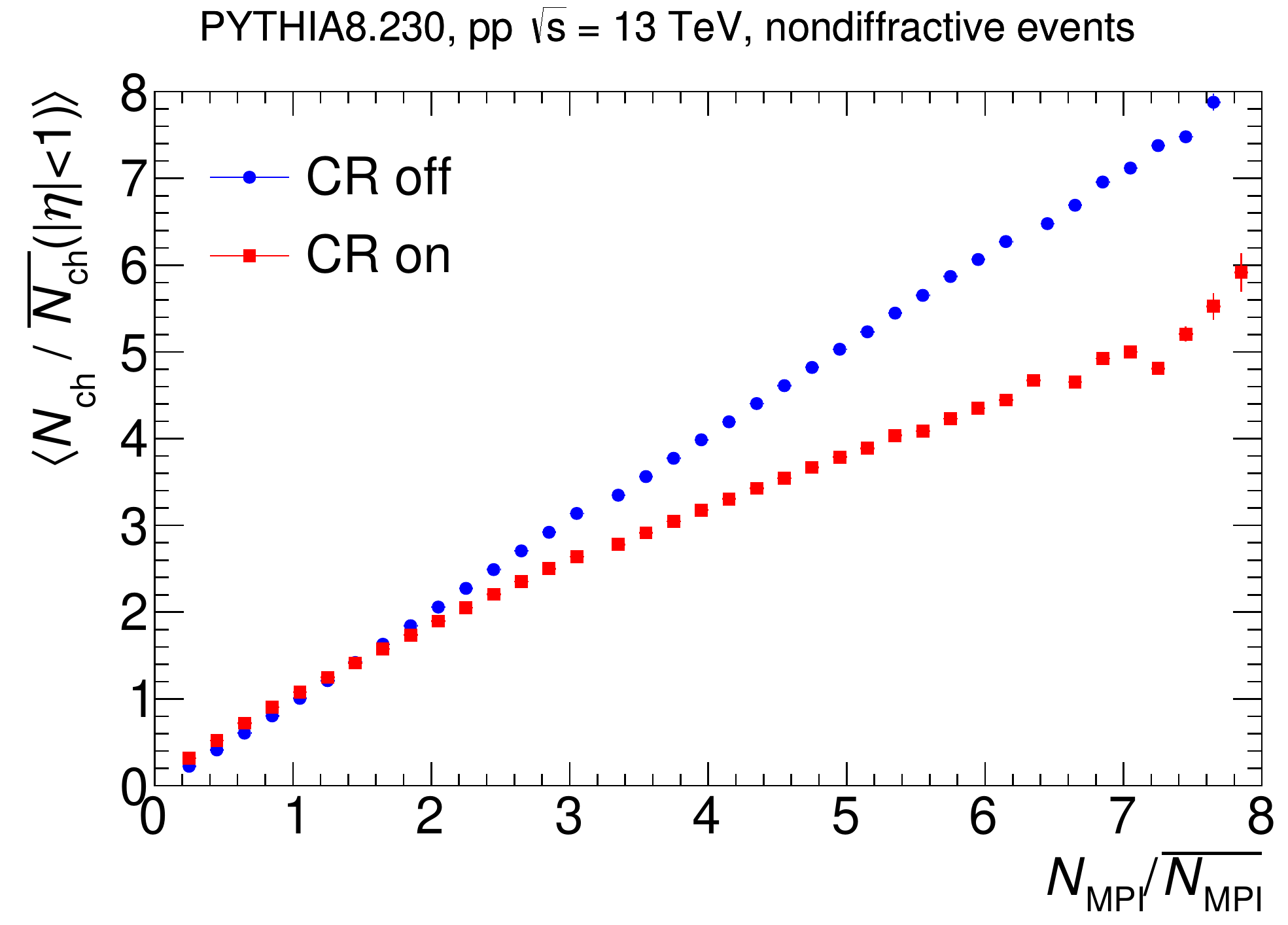}
\end{minipage}
\hspace{0.1cm}
\begin{minipage}[b]{0.5\linewidth}
\centering
    \includegraphics[width=0.9\linewidth]{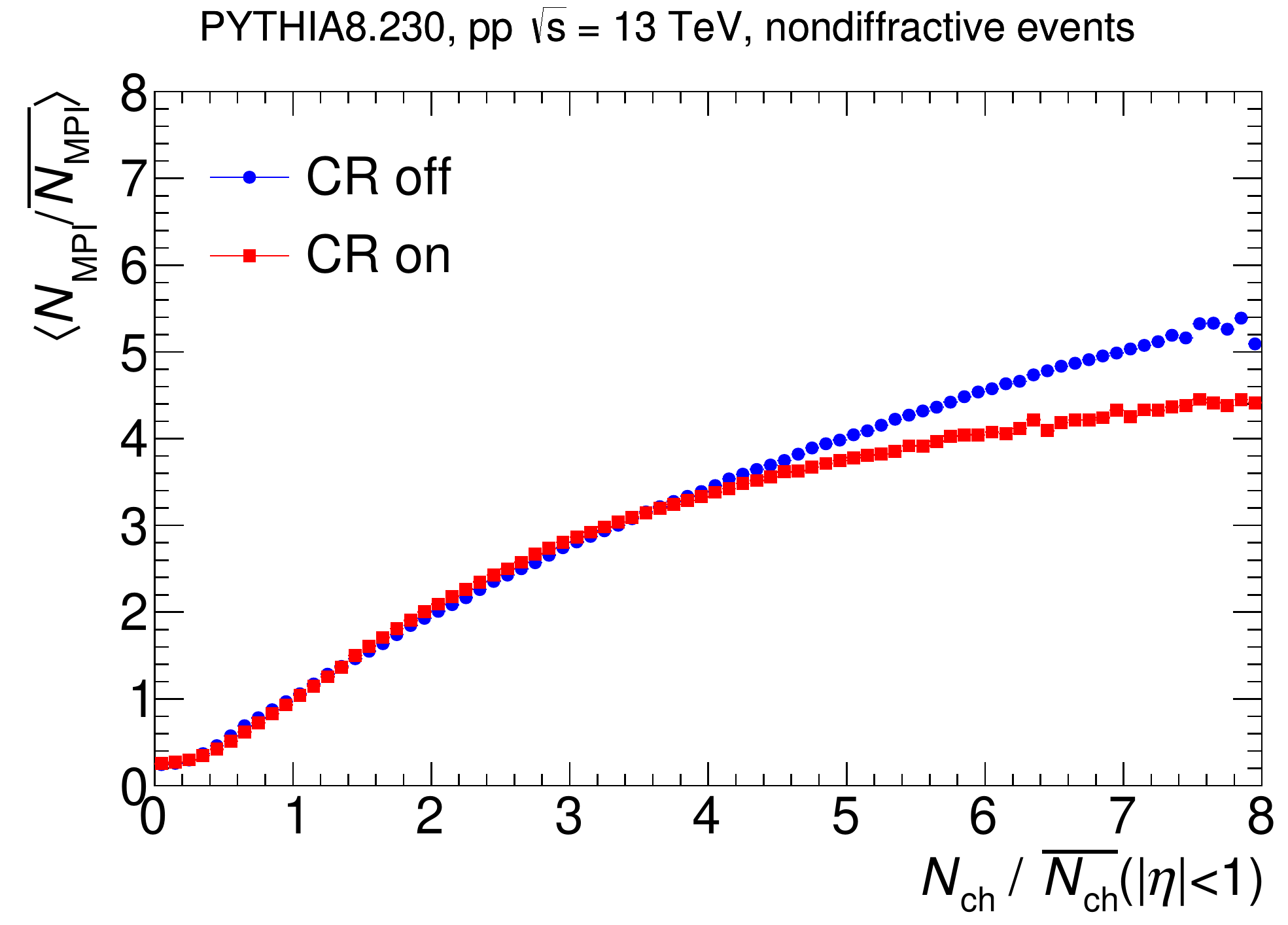}
\end{minipage}%
 \caption{ Left: Mean self-normalized \mult in $|\eta|<1$ as a function of the self-normalized number of MPI for activated and deactivated colour reconnections (CR) in PYTHIA8. Right: Mean self-normalized number of MPI as a function of self-normalized \mult in $|\eta|<1$ in PYTHIA8. }
    \label{fig_pythia_Nch_mpi}
\end{figure*}

\begin{figure*}[ht!]
\begin{minipage}[b]{0.5\linewidth}
\centering
    \includegraphics[width=0.9\linewidth]{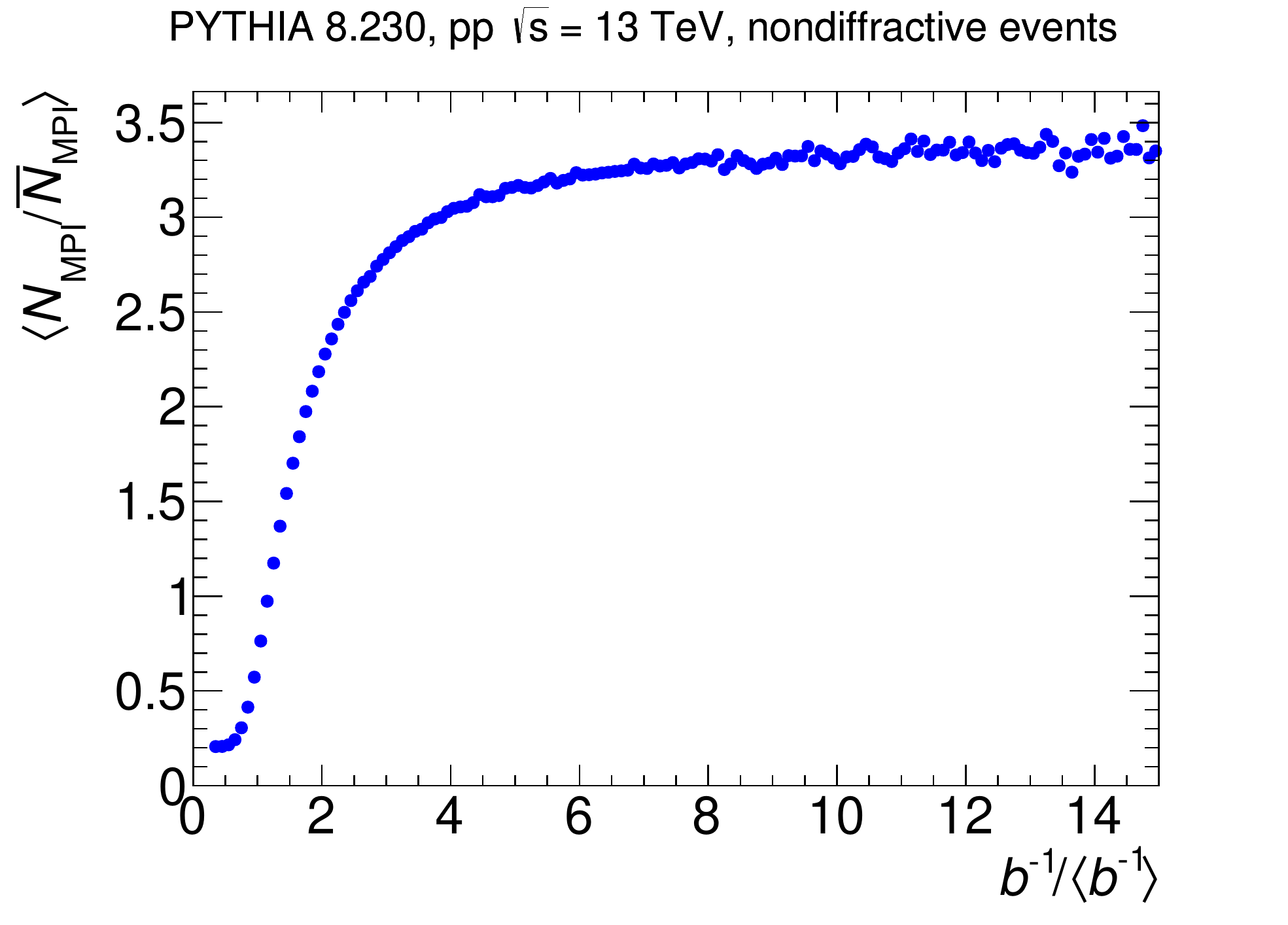}
\end{minipage}
\hspace{0.1cm}
\begin{minipage}[b]{0.5\linewidth}
\centering
    \includegraphics[width=0.83\linewidth]{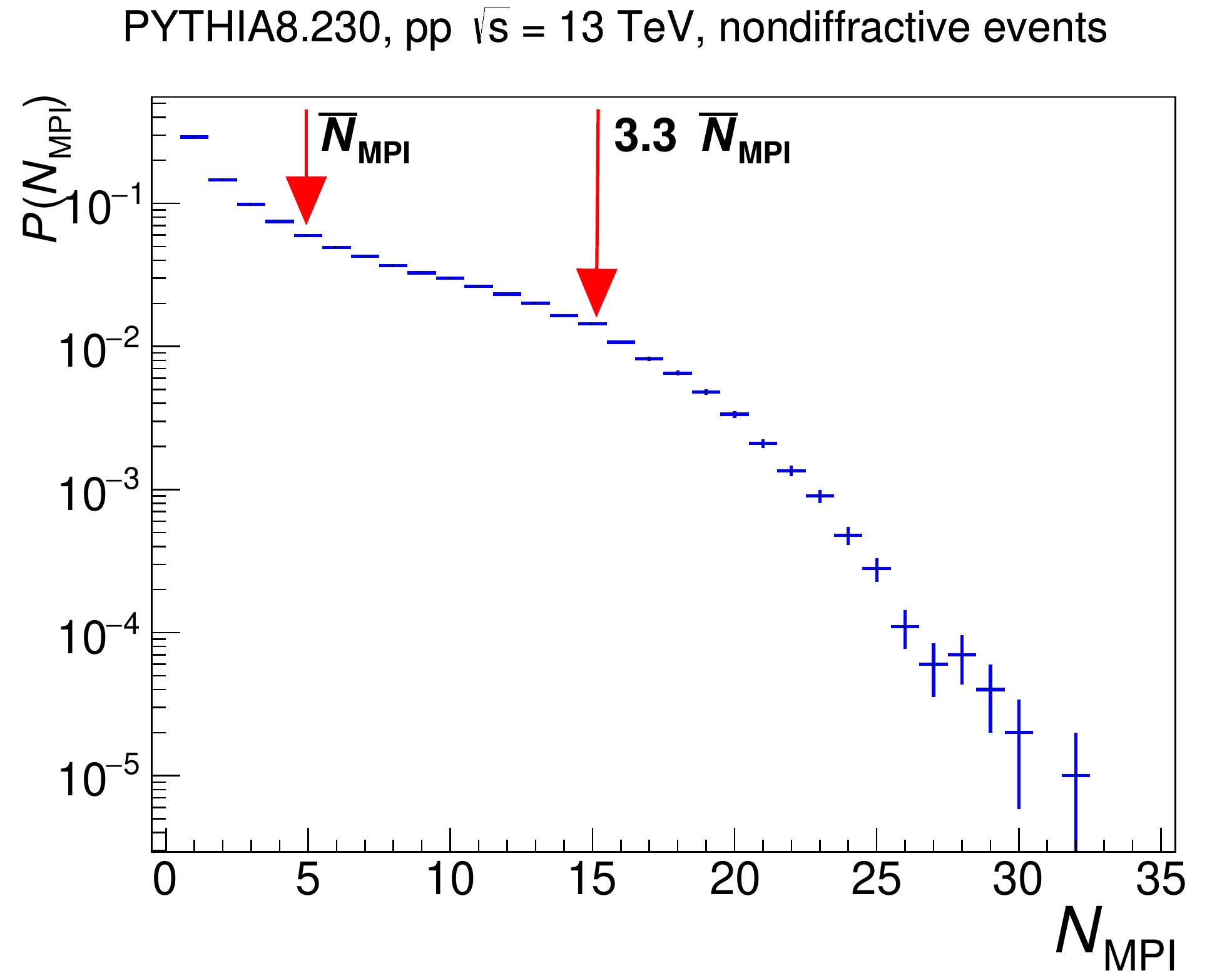}
\end{minipage}%
\caption{Left: Average self-normalized number of MPI per event as a function of the self-normalized inverse impact parameter ($b^{-1}$). Right: The probability distribution of the number of MPI per event. The arrows indicate the average and the 
 change of slope at $3.3 \, \overline{N}_{\rm mpi} $, the maximum value achieved for ($b=0$) without statistical fluctuations. }
    \label{fig_pythia_mpi_b}
\end{figure*}

Without the CR mechanism the individual PIs are independent of each other, thus the \mult \Nch increases roughly linearly with the number of MPI per collision, \Nmpi, as seen in \fig{fig_pythia_Nch_mpi} (left).
The increase is observed to be somewhat weaker than linear, since the total available momentum transfer in one pp collision is fixed, such that for a higher number of MPI the individual PI most likely are softer and produce less particles. With CR switched on, the charged-particle multiplicity distribution for a given \Nmpi gets wider and the mean grows slower as a function of \Nmpi. 
\fig{fig_pythia_Nch_mpi} (right) shows the average \Nmpi as a function of \Nch at mid-rapidity. Since the \Nmpi probability distribution is steeply
falling for large \Nmpi upwards fluctuation of the multiplicity for a given \Nmpi
strongly contribute to high multiplicity events. For this reason the mean \Nmpi grows weaker as linearly with \Nch.
It is interesting to note that CR also reduce the increase of \Nmpi\ with multiplicity. Without the influence of fluctuations one expects the opposite behaviour and for this reason CR have been put 
forward as an explanation for a stronger than linear increase. 
In PYTHIA8, the baseline for the dependence of the yield of hard probes on multiplicity is a 
function (MPI-CR-baseline) which is approximately linear in the range $\Nch / {\overline N}_{\rm ch} \rangle < 3$
and increases weaker than linearly above this value.

Further insight into this deviation from linearity can be obtained by investigating the impact parameter 
dependence of MPI.
As mentioned earlier,  in PYTHIA the number of MPI per event is related to the  
matter overlap in the pp collisions and, hence, to the impact parameter $b$ \cite{pythia_mpi}. 
\fig{fig_pythia_mpi_b} (left panel) shows the average self-normalized number of MPI per event as a function of the self-normalized $b^{-1}$. In the most central collisions, the average number of MPI saturates at 3.3 times the mean value. Even higher number of MPI, as covered in our study, are due to Poissonian fluctuations in the number of MPI towards higher values. In this region the $\Nmpi$ probability distribution is much steeper (see right panel of \fig{fig_pythia_mpi_b}) than at lower values and, hence, the relation between $\Nmpi$ and $\Nch$ is more sensitive to upwards fluctuations of 
the multiplicity produced by individual PI.

\section{Heavy quark and quarkonium production in PYTHIA8}
\label{sec_pythia_heavy}

\begin{figure*}[ht!]
\begin{minipage}[b]{0.5\linewidth}
\centering
    \includegraphics[width=0.9\linewidth]{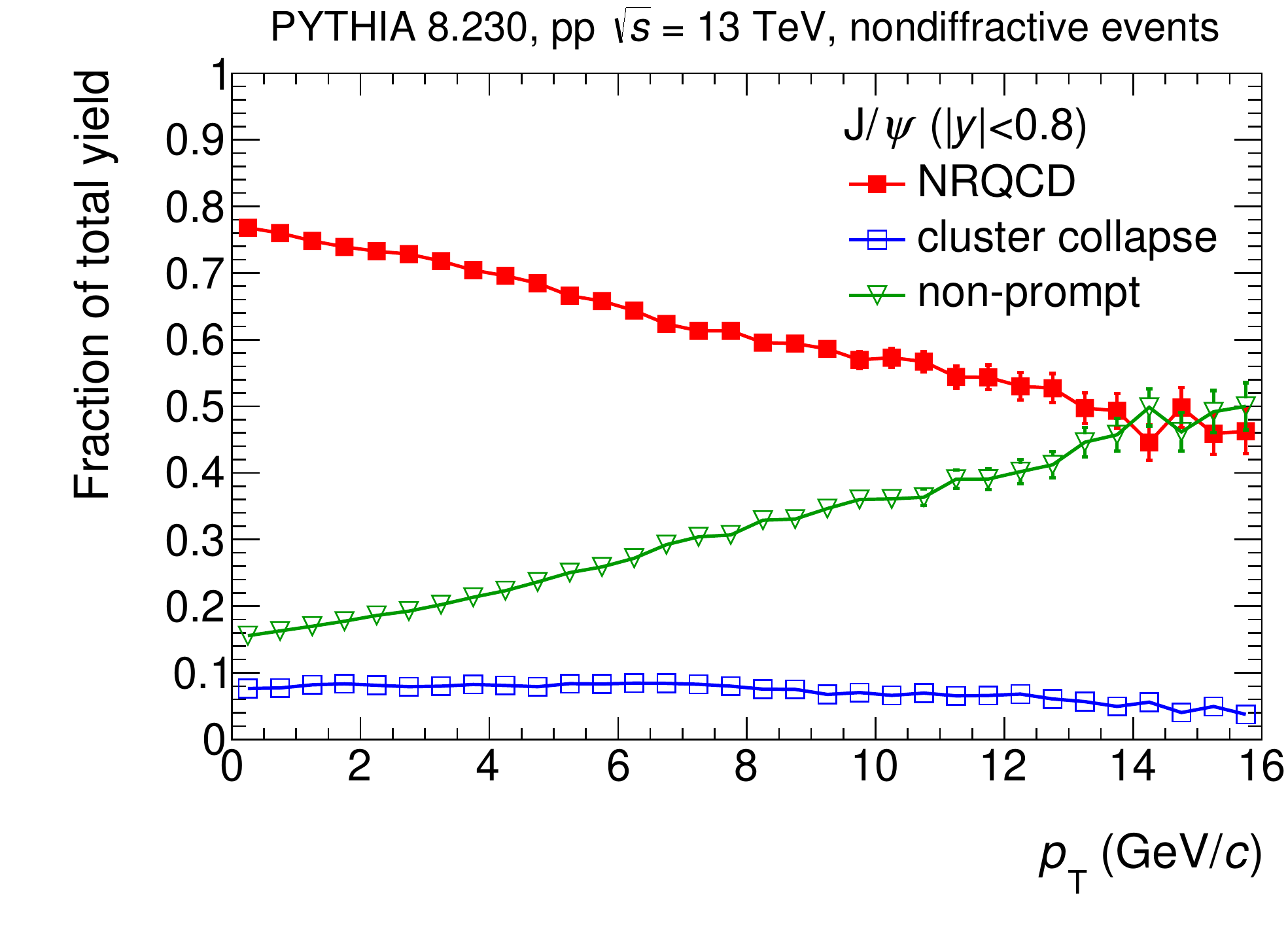} 
\end{minipage}
\hspace{0.1cm}
\begin{minipage}[b]{0.5\linewidth}
\centering
        \includegraphics[width=0.9\linewidth]{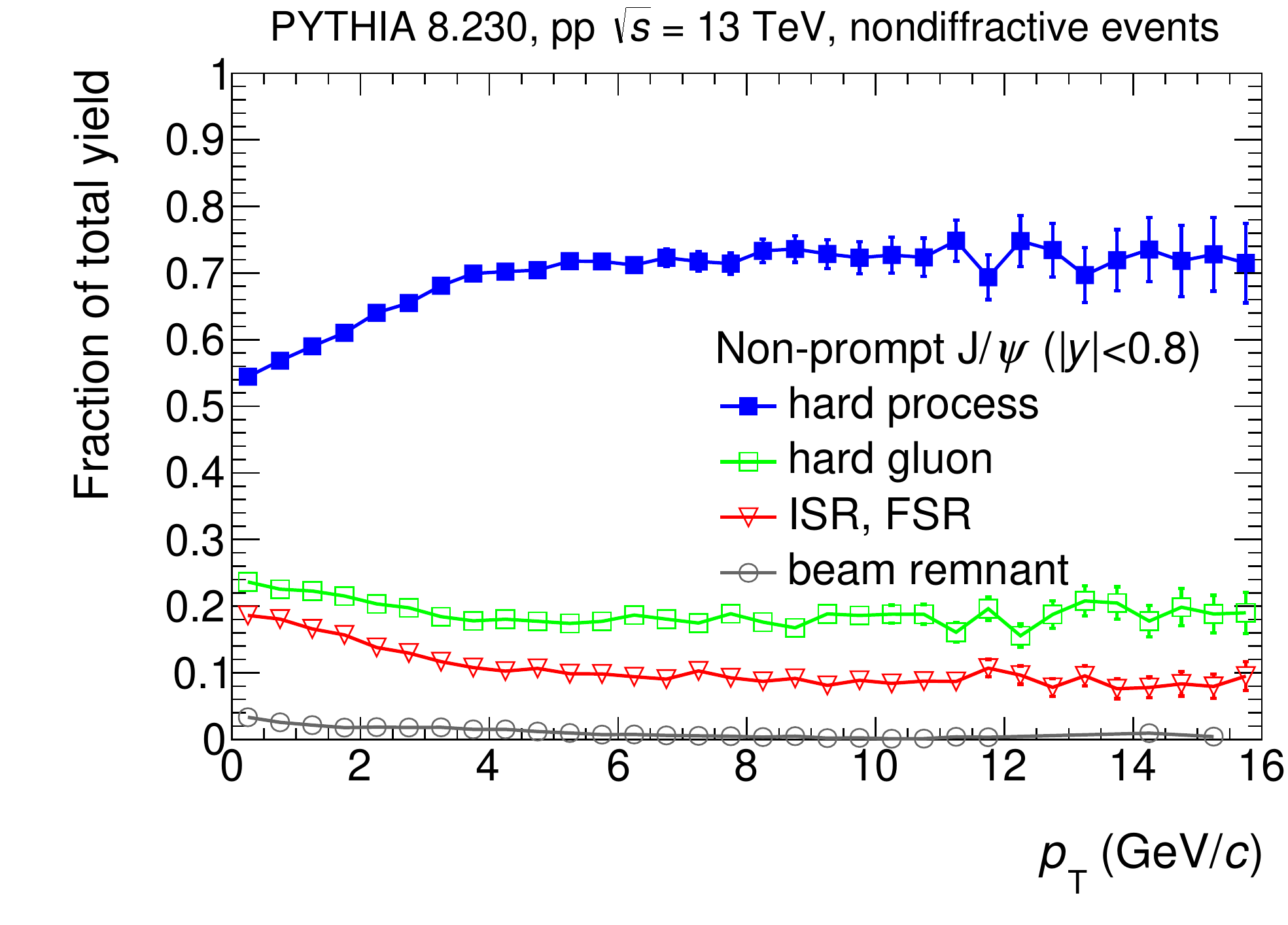} 
\end{minipage}%
 \caption{Left: Relative contributions of different production processes to mid-rapidity \jpsi production as a function of transverse momentum in PYTHIA8. Right: Relative contributions of the different heavy quark production mechanism to the non-prompt \jpsi yield as a function of transverse momentum in PYTHIA8. }
    \label{fig_pythia_kinematics_sources}
\end{figure*}

Heavy quark pair production is implemented in PYTHIA as perturbative scattering processes in the form of gluon fusion $gg\rightarrow Q\bar{Q}$ ($Q$ denotes either charm or bottom) or light quark-antiquark annihilation $q\bar{q}\rightarrow Q\bar{Q}$.
At the $Q^2$ of the hard scattering, heavy quarks can be also present in the parton distribution function leading to heavy quark production via 
$Qg \rightarrow Qg$ (flavour excitation). Moreover, in the parton shower heavy quarks can be produced by gluon splitting $g \rightarrow Q\bar{Q}$.

For quarkonia, different production mechanisms are implemented. First, in the perturbative scattering processes, leading order NRQCD channels via colour-singlet and colour-octet pre-resonant states are included \cite{pythia82_onia}.
For colour-octet states, one additional gluon is emitted in the transition to the physical colour-singlet state.
Secondly, quarkonia can be produced from the cluster collapse mechanism \cite{charm_string}. This occurs when at the hadronization stage, a heavy quark is connected to a corresponding heavy antiquark. If they are close in phase space the potential energy in the string is insufficient to create a light quark-antiquark pair, and instead the two heavy quarks bind into a quarkonium bound state.
Finally, in so called non-prompt charmonium production,  charmonia can be produced from the weak decay of a hadron containing a beauty quark.

In the left panel of \fig{fig_pythia_kinematics_sources} the relative contributions of the different sources of \jpsi production to the total \jpsi yield at mid-rapidity are shown as a function of \pt . 
The largest fraction of \jpsi is produced in NRQCD processes. Non-prompt \jpsi production has a harder \pt spectrum than prompt \jpsi and its relative contribution rises from about $\sim$10\% at \pt = 0 to above $\sim$40\% at \pt $\sim$16 GeV/$c$. These observations are in agreement with experiment, see e.g. \cite{alice_nonprompt}.
The cluster collapse contribution amounts to between $\sim$8\% and $\sim$4\% of the total \jpsi yield, depending on \pt. 

For non-prompt \jpsi the origin of the initial beauty quark is investigated. This is shown in the right panel of \fig{fig_pythia_kinematics_sources}, which depicts the relative contributions of the different sources of beauty quark production to the total yield of non-prompt \jpsi at mid-rapidity as a function \pt.
About $\sim$65\% of all beauty quarks are produced in a primary perturbative scattering process, about $\sim$20\% in the splitting of a gluon, which was in turn produced in a hard scattering, and the remaining $\sim$15\% in the splitting of a gluon from initial or final-state radiation. In the latter case, the \pt distribution is slightly softer than for leading order processes. The contribution of beauty quarks from beam remnants is negligible at mid-rapidity.

\section{Results}

In our MC simulations, the \jpsi was forced to decay in the dielectron channel in order to have the same conditions as in experiments that typically reconstruct \jpsi either from this or the dimuon decay channel.
The role of MPI, CR and auto-correlations, are investigated in the following in order to understand the origin of the experimentally observed stronger than the MPI-CR-baseline increase and the \pt dependence \cite{preliminary}.

\subsection{Multi-Parton Interactions and Colour Reconnections}

\begin{figure*}[ht!]
\begin{minipage}[b]{0.5\linewidth}
\centering
    \includegraphics[width=0.9\linewidth]{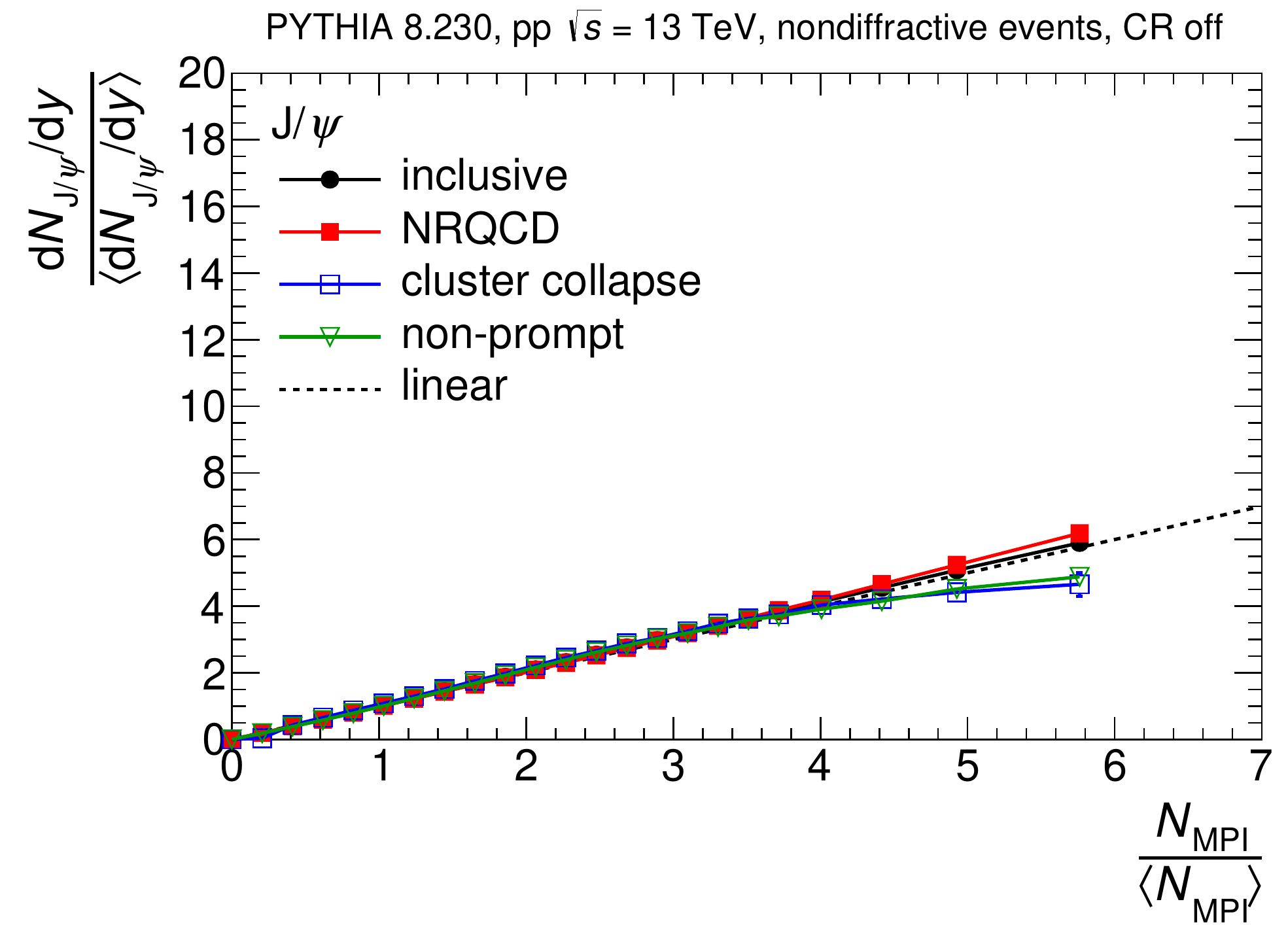} 
\end{minipage}
\hspace{0.1cm}
\begin{minipage}[b]{0.5\linewidth}
\centering
    \includegraphics[width=0.9\linewidth]{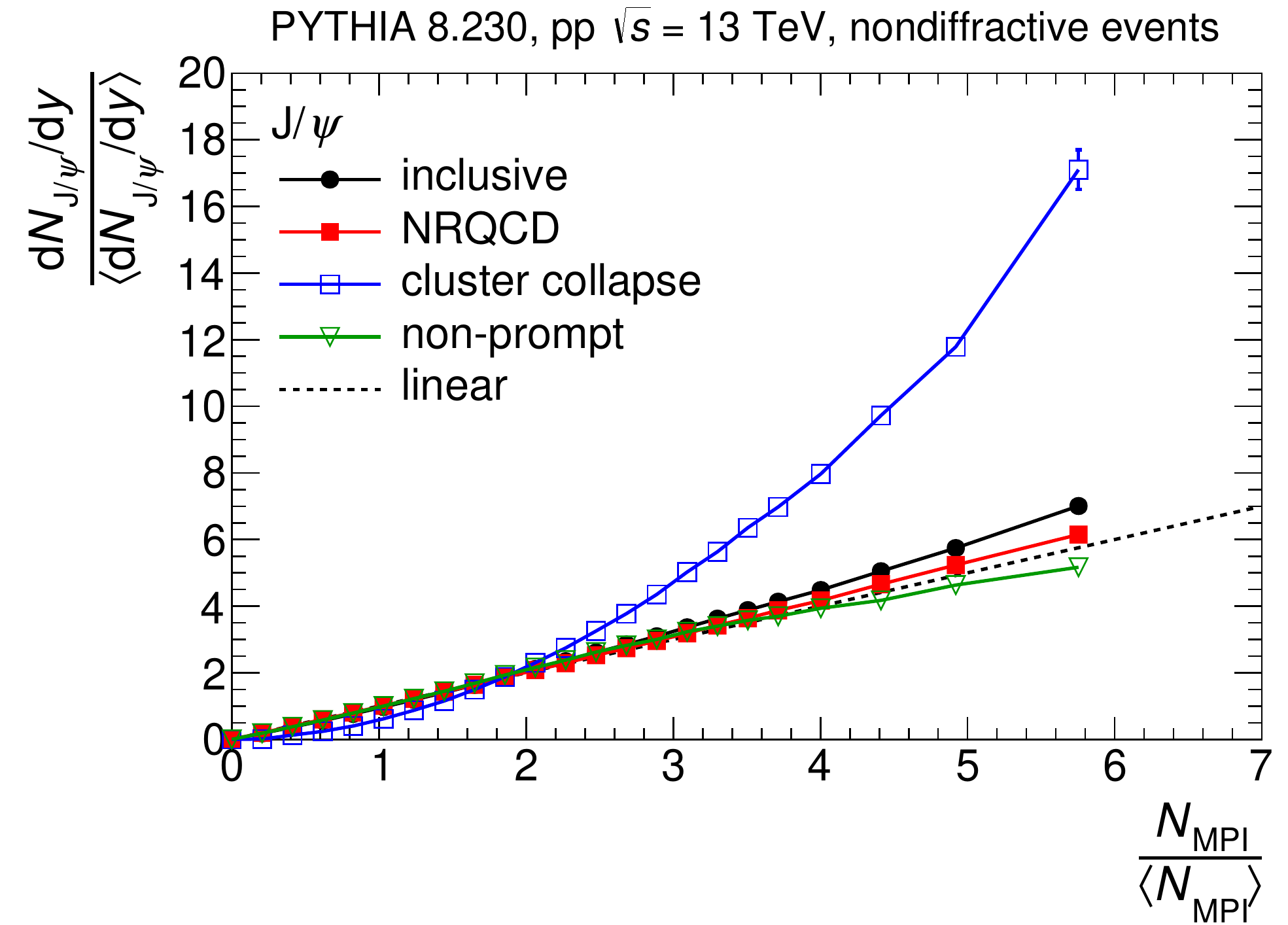} 
\end{minipage}%
  \caption{Self-normalized \jpsi yield from non-diffractive events, generated with PYTHIA8, split in different production processes as a function of the number of \mpi. Left: colour reconnection switched off, right:  colour reconnections on.}
    \label{fig_pythia_panels_sources_mpi}
\end{figure*}

\noindent In the PYTHIA model each PI has a certain probability to produce a \jpsi. Consequently the self-normalized \jpsi yield rises approximately linearly with the self-normalized number of MPI, as shown in the left panel of \fig{fig_pythia_panels_sources_mpi}.

\begin{figure*}[ht!]
\begin{minipage}[b]{0.5\linewidth}
\centering
    \includegraphics[width=0.9\linewidth]{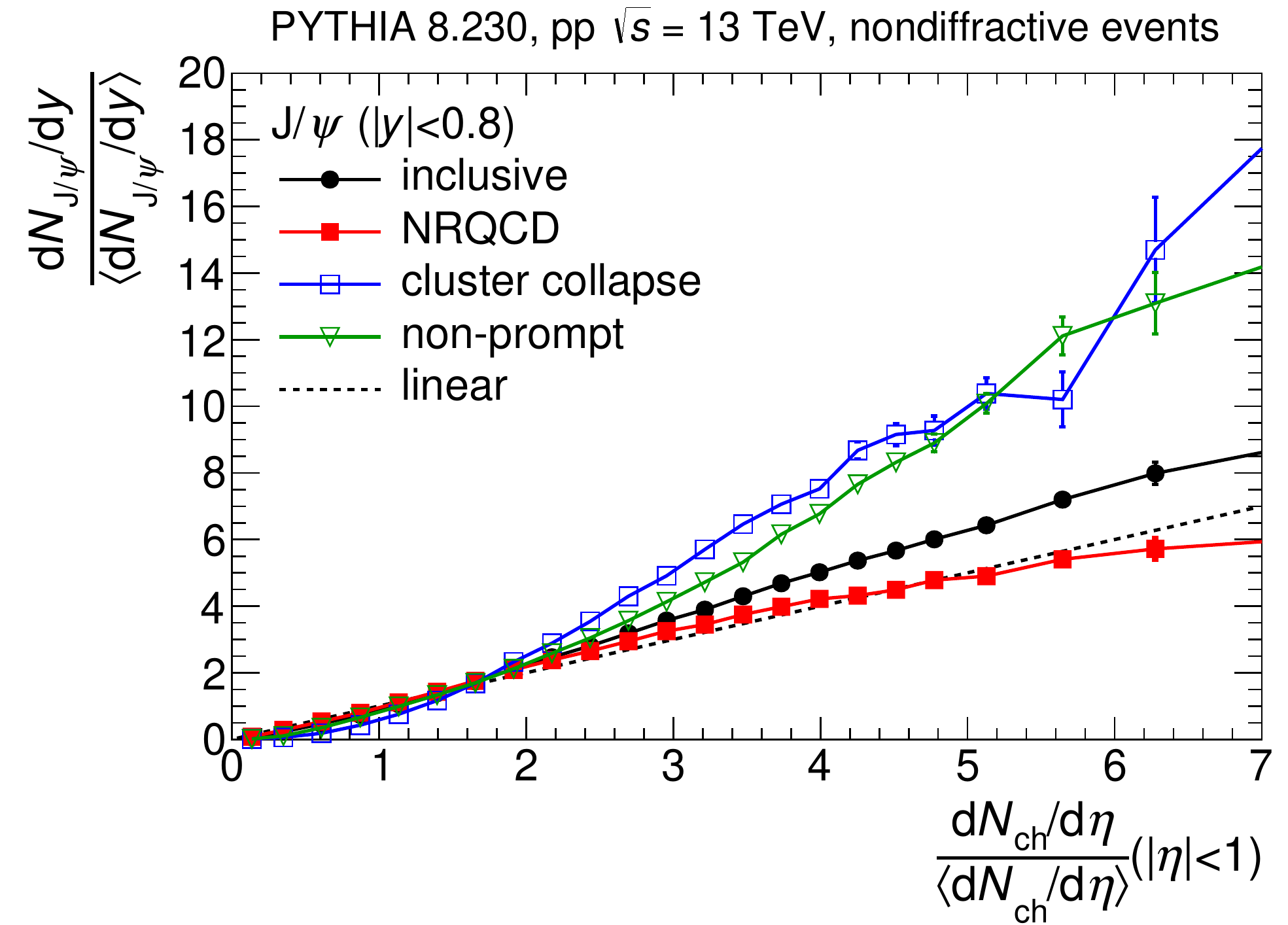} 
\end{minipage}
\hspace{0.1cm}
\begin{minipage}[b]{0.5\linewidth}
\centering
    \includegraphics[width=0.9\linewidth]{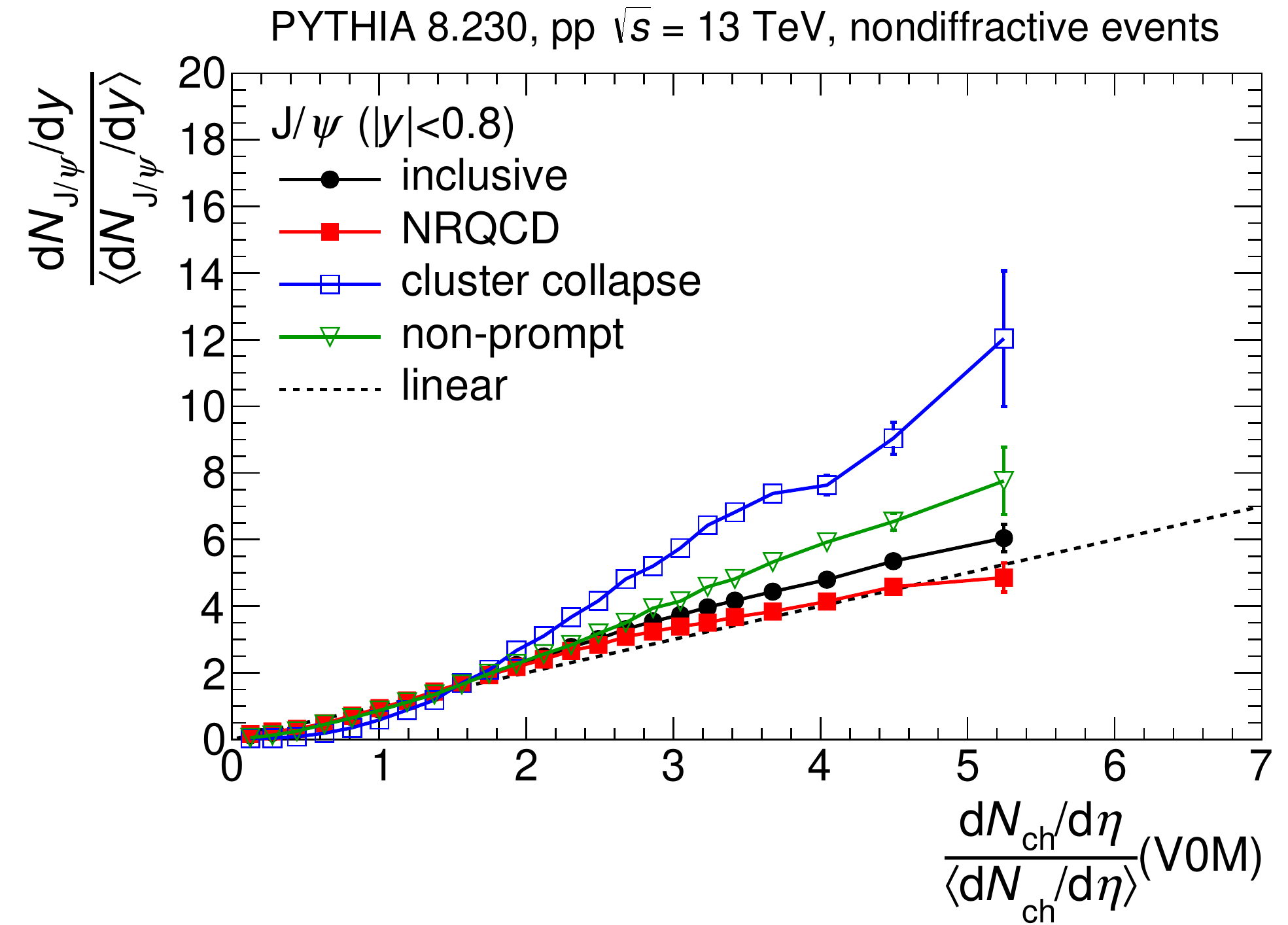} 
\end{minipage}%
 \caption{ Mid-rapidity \jpsi production as a function of \mult at mid (left) and forward (right) rapidity split into the different production processes implemented in PYTHIA8, i.e. non-prompt \jpsi, prompt \jpsi from NRQCD and prompt \jpsi from cluster collapse. }
    \label{fig_pythia_sources_yield}
\end{figure*}

Here CR was deactivated, as its influence will be discussed later.
In this scenario, the \jpsi yields from NRQCD vs \Nmpi is closest to linear. For non-prompt \jpsi and \jpsi yields from cluster collapse, the increase tends to saturate at very high multiplicities. This can be explained by the fact that the total available energy in one collision is limited. Hence, for a large number of PIs in one collision, the momentum transfer per single PI is on average smaller, so the cross section for hard processes such as heavy quark production is reduced.

The right panel of \fig{fig_pythia_panels_sources_mpi} shows the same dependence with CR activated. \jpsi production from NRQCD and non-prompt \jpsi are not affected by CR. This is as expected, since CR acts on the colour strings that are responsible for light particle production but does not enter in these \jpsi production processes. On the other hand \jpsi from cluster collapse shows a completely different behaviour, as the yield grows quadratically with \Nmpi. In this process, the charm and the anti-charm quark typically originate from independent pairs, since heavy quarks produced as a pair usually have a large opening angle. 
With CR activated, also a charm quark from one PI can bind with an anticharm quark from a different PI. The probability to produce one heavy quark pair in a collision increases linearly with \Nmpi: $P(c\bar c) \propto \Nmpi$. Since the processes are independent, the probability to produce a second one under the condition that a first one was produced also increases linearly with \Nmpi: $P( 2 c\bar c | c\bar c ) \propto \Nmpi$, so the total probability to produce two charm-anticharm pairs increases quadratically  $P( 2 c\bar c ) = P( 2 c\bar c | c\bar c ) \cdot P(c\bar c) \propto \Nmpi^2 $.

\begin{figure*}[ht!]
\begin{minipage}[b]{0.5\linewidth}
\centering
   \includegraphics[width=0.9\linewidth]{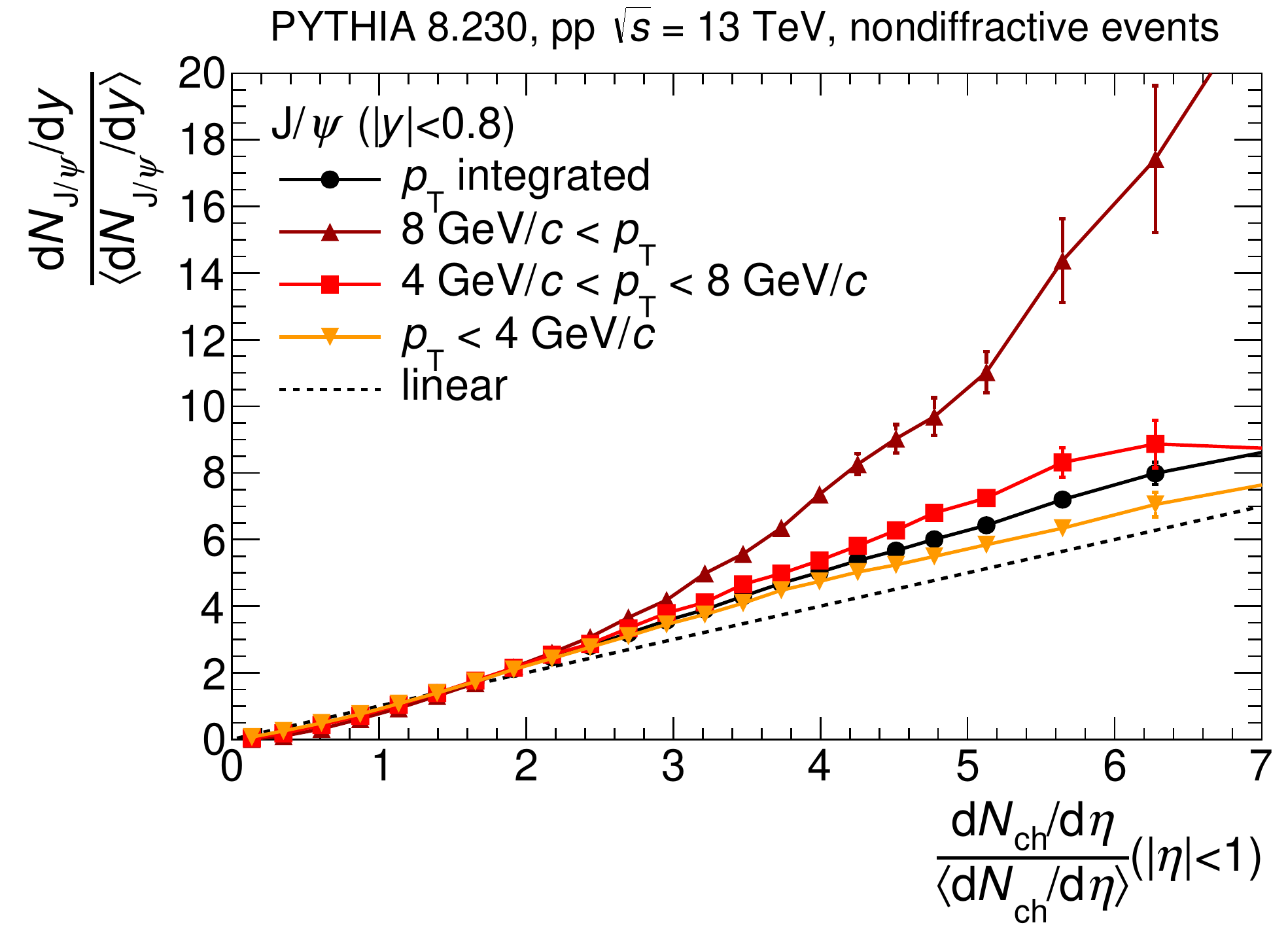}
\end{minipage}
\hspace{0.1cm}
\begin{minipage}[b]{0.5\linewidth}
\centering
    \includegraphics[width=0.9\linewidth]{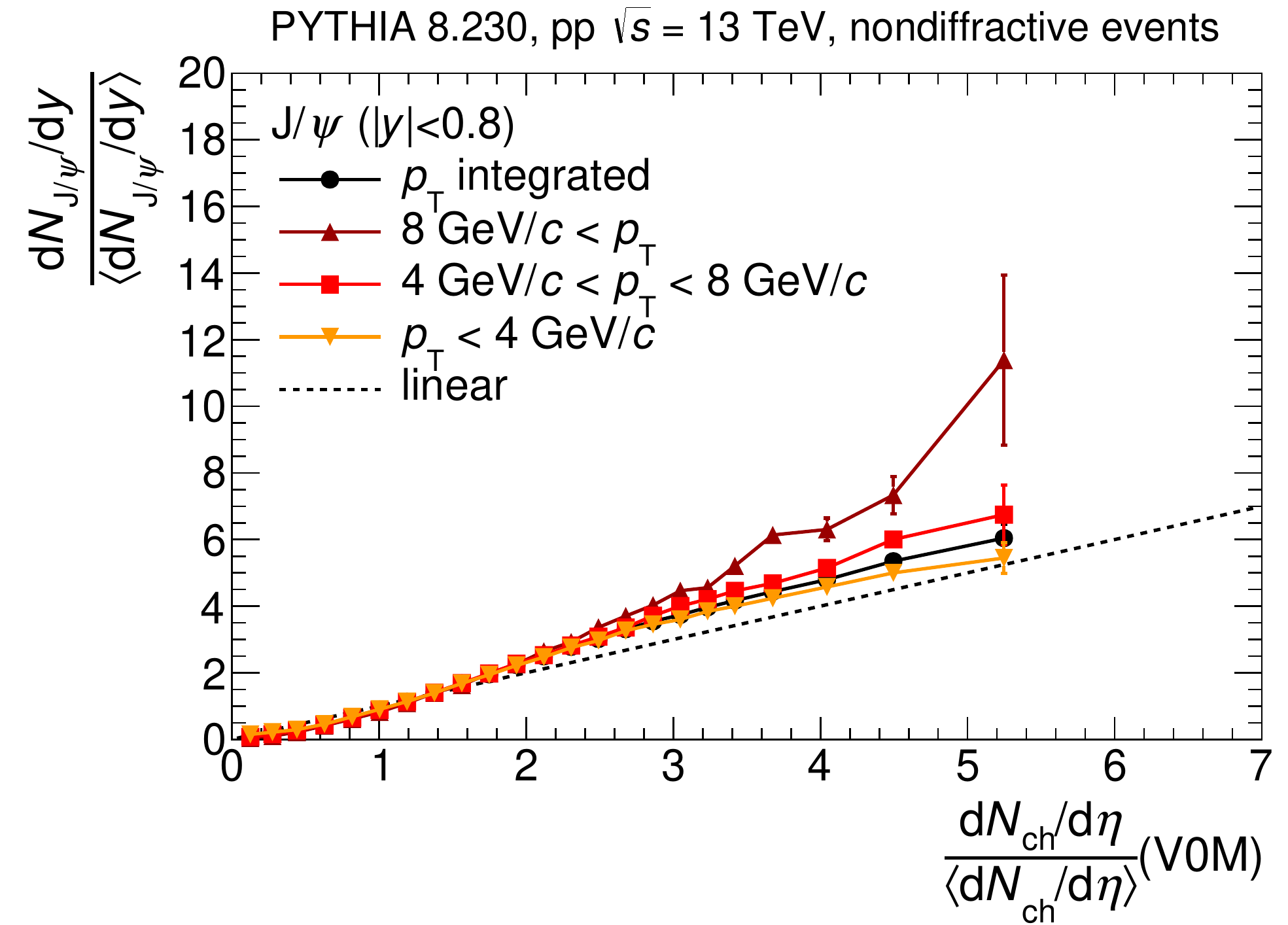} 
\end{minipage}%
 \caption{ Mid-rapidity \jpsi production as a function of \mult at mid (left) and forward (right) rapidity in different \pt intervals. }
    \label{fig_pythia_pt_yield}
\end{figure*}

In \fig{fig_pythia_sources_yield} the self-normalized yields of \jpsi, for the different production mechanism, are shown as a function of the charged-particle multiplicity estimated at mid-rapidity (left panel) and forward rapidity (right panel). The \jpsi yield from NRQCD grows linearly with multiplicity, independent of whether the latter was measured at mid or forward rapidity. The \jpsi yield from cluster collapse grows stronger than linearly independently from the rapidity where the multiplicity was measured. Also the non-prompt \jpsi yield grows stronger than linearly, however, in this case the increase is stronger as a function of the mid-rapidity than the forward rapidity multiplicity. This observation hints to the possibility that for non-prompt \jpsi  auto-correlation effects are important. 
Note that also the linear behaviour of the \jpsi yield from NRQCD lies above the MPI-CR-baseline and,
hence, also in this case auto-correlation effects might have an influence.
The \pt dependence of the increase is shown in \fig{fig_pythia_pt_yield}. Both as a function of mid-rapidity (left) and forward-rapidity (right) multiplicity, the increase is steeper with rising \pt, in agreement with what has been experimentally measured \cite{preliminary}. The effect is slightly stronger for the mid-rapidity multiplicity and it is mostly due to the non-prompt \jpsi contribution, where the dependence on \pt is most pronounced.

\subsection{Auto-correlation effects}

\noindent The \mult and the \jpsi yield are not independent quantities, the latter influences the former. This influence comes from the following mechanisms: \\

\begin{itemize}
\item The \jpsi decay daughters enter the \mult. The simulations, in order to have the same feature of the experimental measurement are done in the dielectron decay channel. In this case two additional charged particles are produced in events containing a \jpsi, if the multiplicity is measured in the same rapidity as the \jpsi. \\

\item In NRQCD processes the \jpsi is typically produced together with a gluon, e.g. via $g\,g \rightarrow [Q\bar{Q}]\,g$ which will in turn hadronize and increase the multiplicity. Additionally, if the pre-resonant state is a colour-octet, an additional gluon is emitted in the transition to the physical \jpsi state. Since the mass difference between the colour-octet and the colour-singlet state is small, the gluon will typically be emitted under a small opening angle, so the multiplicity in the flight direction of the \jpsi will be affected most. \\

\item In the case of \jpsi from cluster collapse, the charm quark and antiquark are both produced together with another charm antiquark and charm quark, which in turn will produce additional particles. \\

\item In the case of non-prompt \jpsi, the mother particle of the \jpsi decays into several particles and the decay daughters can decay further. Furthermore, the initially produced beauty quark can be accompanied by final state radiation, enhancing the multiplicity in the region around it.
Finally, beauty quarks are always produced as pairs, mostly back-to-back in hard interactions. Thus, the non-prompt \jpsi will typically be accompanied by a high \pt parton going in the opposite direction, fragmenting into a jet of particles. This recoil jet is at an azimuthal angle of $180^\circ$ with respect to the initial b quark, but can be at a different rapidity.
\end{itemize}

\noindent These auto-correlation effects can be best studied in events with only one hard interaction, which is with the MPI mechanism switched off. Then, \jpsi and charged particle production both originate from the same process and the described effects should be clearly visible.

\begin{figure}[th!]
  \centering
    \includegraphics[width=0.95\linewidth]{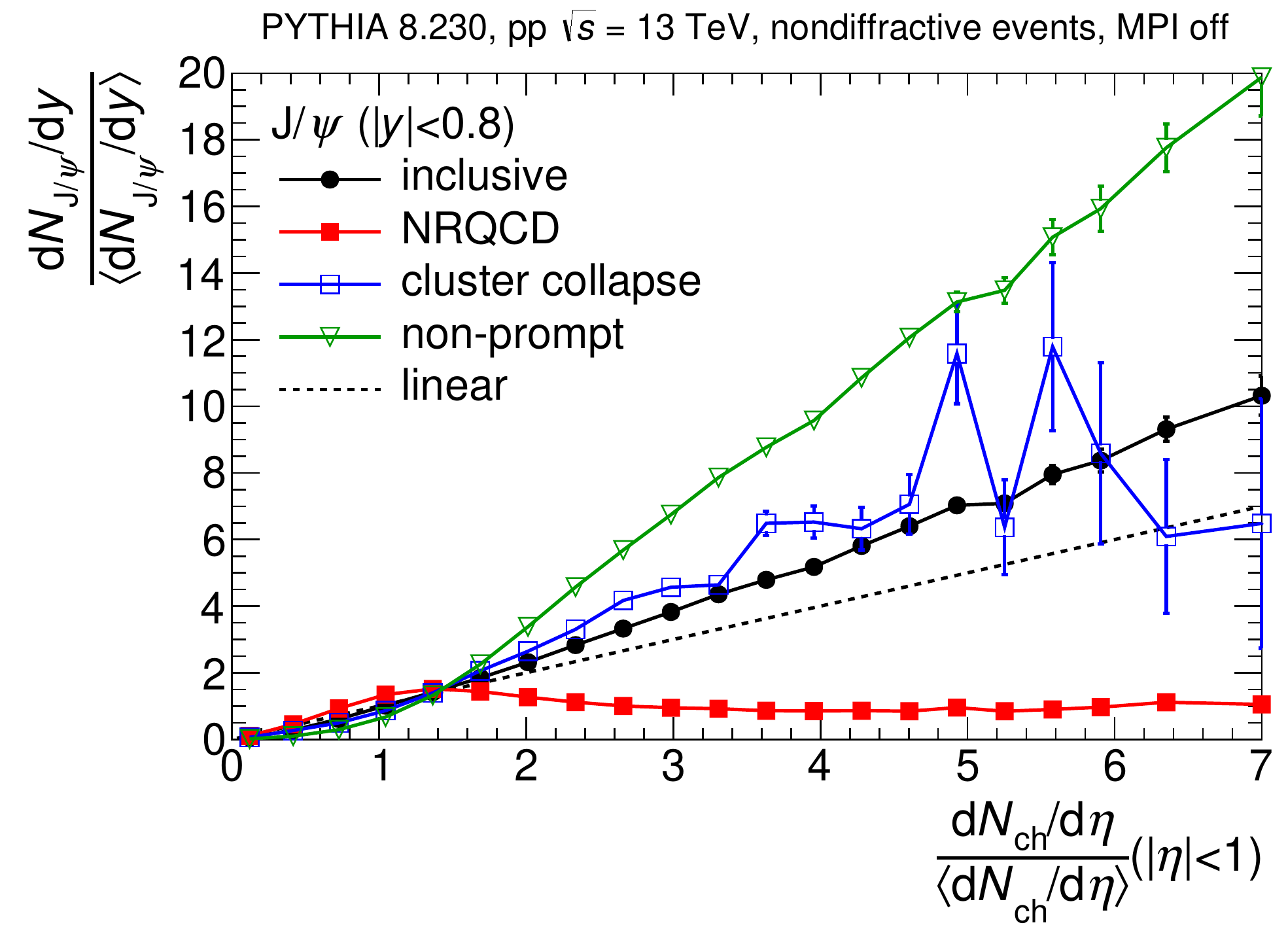} 
  \caption{ Mid-rapidity \jpsi production as a function of mid-rapidity multiplicity in events without MPI from PYTHIA8.}
    \label{fig_pythia_noMPI}
\end{figure}

\begin{figure}[th!]
  \centering
    \includegraphics[width=0.45\linewidth]{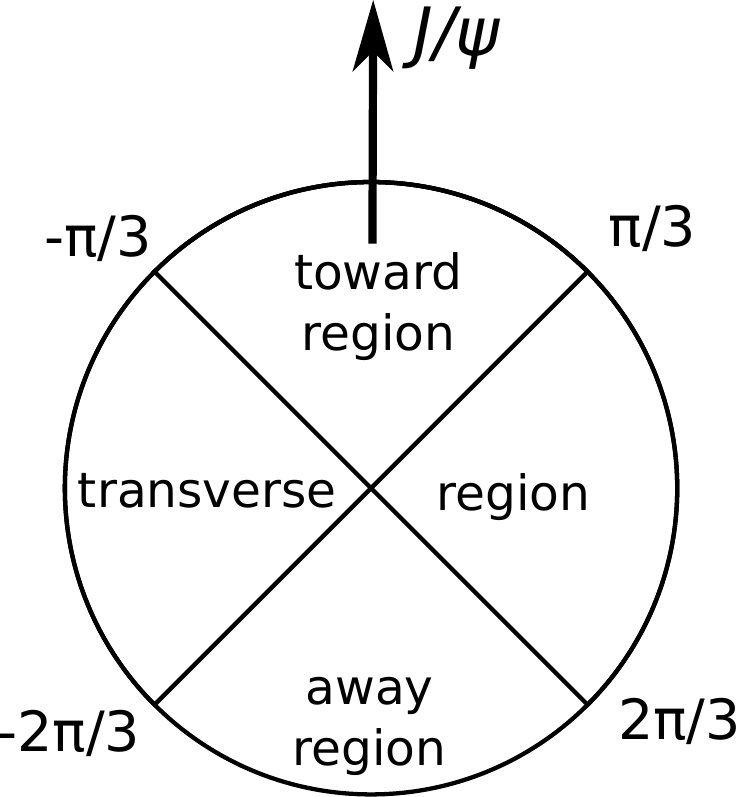} 
  \caption{Definition of the Toward, Transverse and Away region in $\varphi$ with respect to the \jpsi direction. }
    \label{fig_pythia_regions_definition}
\end{figure}

\begin{figure*}[th!]
  \centering
    \includegraphics[width=0.75\linewidth]{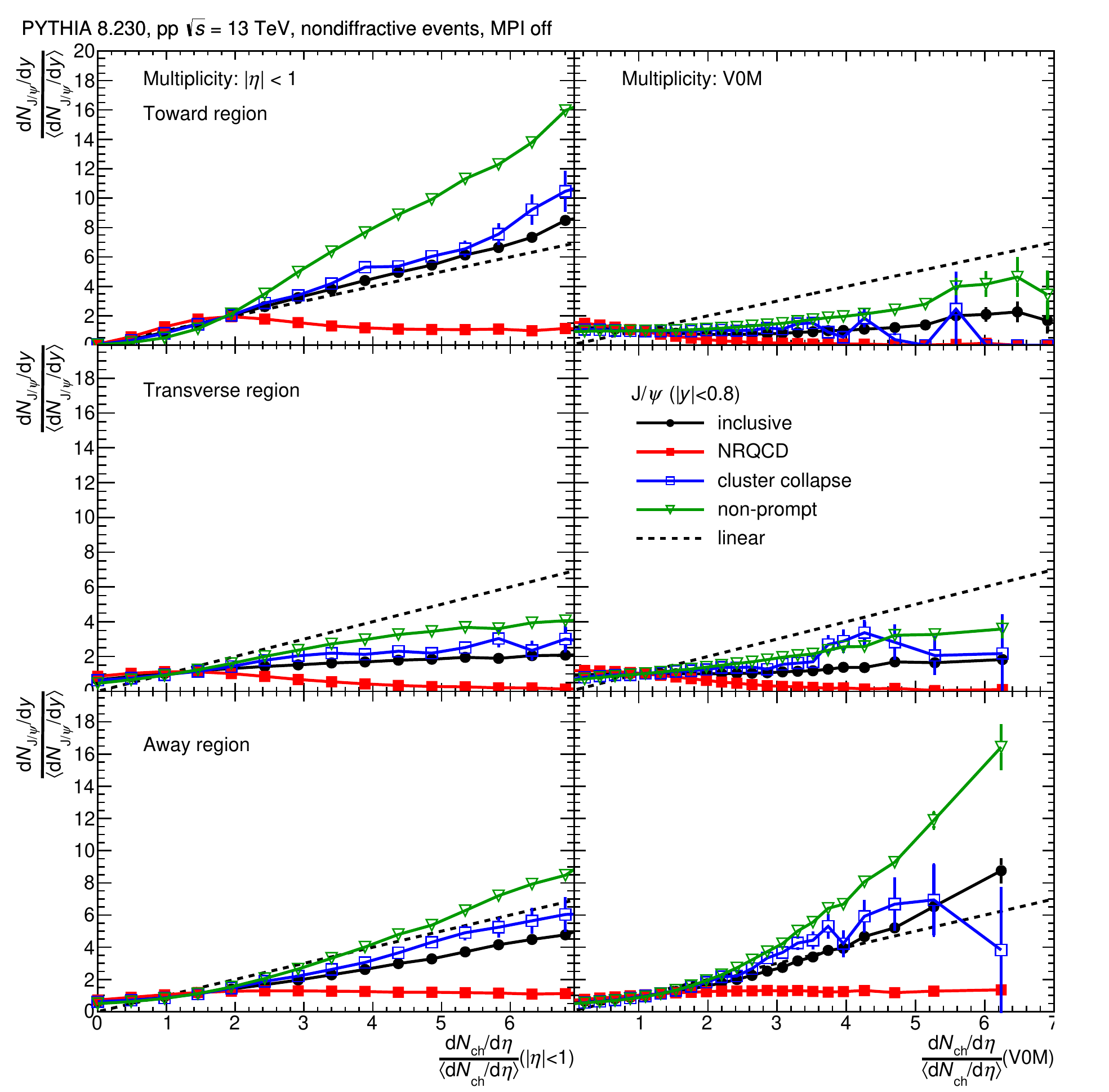} 
  \caption{Multiplicity dependence of \jpsi produced at mid-rapidity in PYTHIA8.230 with \mpi switched off, split into the different production processes. The different panels are for the multiplicity evaluated in different kinematic regions. Left: multiplicity at mid-rapidity, right: multiplicity at forward rapidity; top row: multiplicity in Toward region in $\varphi$ w.r.t. \jpsi, middle row: multiplicity in Transverse region, bottom row: multiplicity in Away region.}
    \label{fig_pythia_noMPI_regions_sources}
\end{figure*}

\fig{fig_pythia_noMPI} shows the self-normalized \jpsi yield at mid-rapidity as a function of the self-normalized \mult at mid-rapidity for events without MPI. For non-prompt \jpsi, a strong increase of the yield with the \mult can be observed, likewise but weaker for \jpsi from cluster collapse. For \jpsi from NRQCD a different picture emerges: at low multiplicity, the yield increases with multiplicity, up to around 1.5 times the mean value, afterwards it decreases again slightly with multiplicity. This behavior can be understood in the following way: at low multiplicity the additional particle production from the gluons produced alongside the \jpsi leads to an increase of \jpsi production with multiplicity. At higher multiplicity, the competition between \jpsi and charged particle production for the limited total phase space in the collision becomes relevant and leads to the observed decrease of the self-normalized \jpsi yield.

To further investigate the different auto-correlation effects, the \jpsi yield can be studied as a function of the \mult in different angular regions with respect to the direction of the \jpsi, i.e. in different regions of the azimuthal angle $\varphi$ and at different rapidities. The $\varphi$ direction is split into three regions as indicated in \fig{fig_pythia_regions_definition}:
\begin{itemize}
\item Toward region: $\Delta \varphi \equiv |\varphi-\varphi_{\jpsi} |< \pi /3$\\
\item Transverse region: $ \pi/3 < \Delta \varphi  < 2 \pi /3$\\
\item Away region: $ 2 \pi /3 < \Delta \varphi$.
\end{itemize}
\noindent Furthermore, the \mult was determined either at mid-rapidity, or at forward rapidity.
The dependence of mid-rapidity \jpsi production on the multiplicity in the different regions is shown in \fig{fig_pythia_noMPI_regions_sources}. In the left panels, the multiplicity was determined at mid-rapidity, in the right ones at forward rapidity; the top, middle, and bottom panels are for the multiplicity determined in the Toward, Transverse, and Away regions, respectively.

\begin{figure*}[th!]
  \centering
    \includegraphics[width=0.75\linewidth]{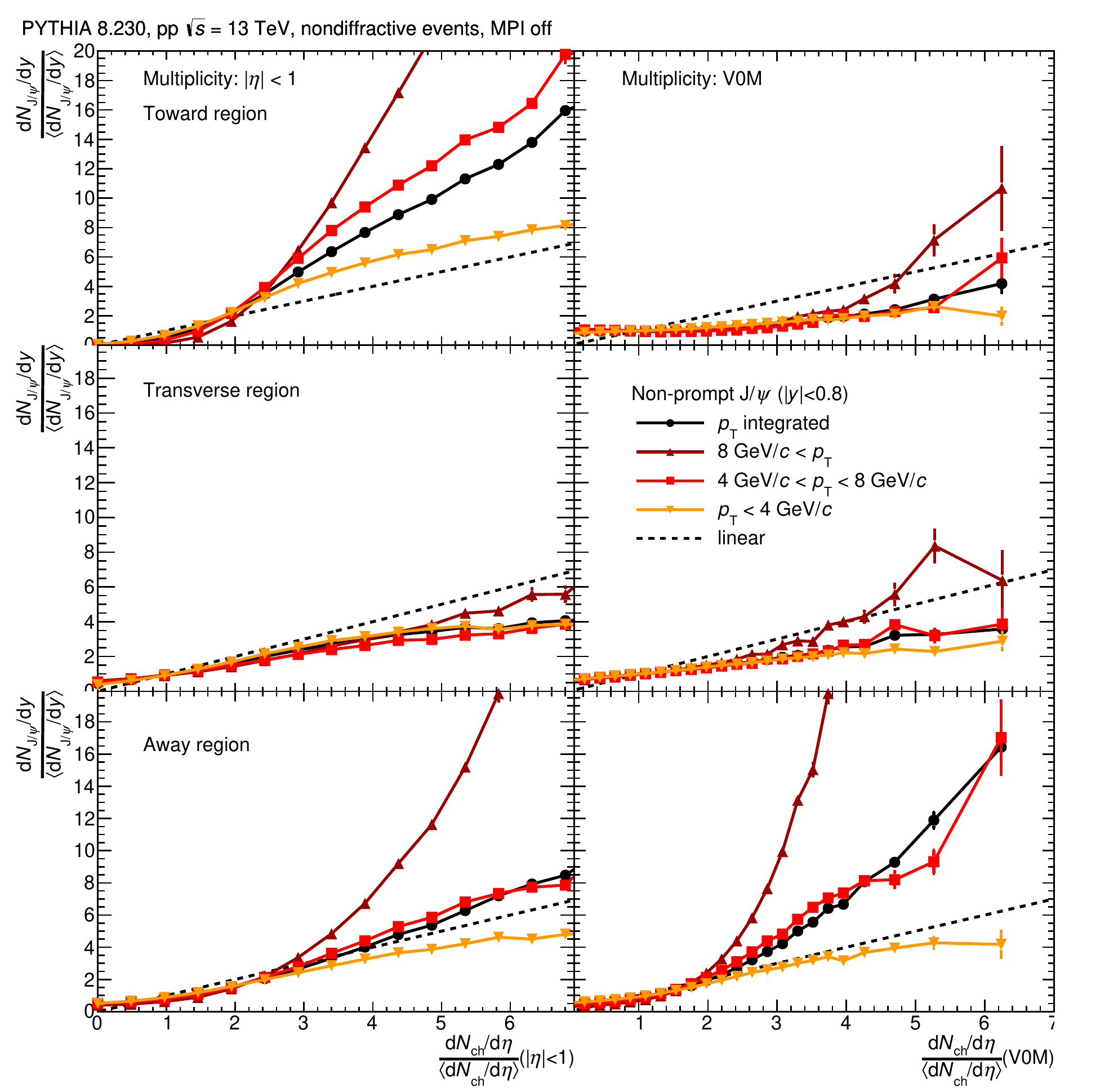} 
  \caption{Multiplicity dependence of non-prompt \jpsi in different \pt intervals produced at mid-rapidity in PYTHIA8.230 with \mpi switched off, for the beauty quark produced in a hard pQCD process. 
  The panels are as in Fig.~\ref{fig_pythia_noMPI_regions_sources}.
  }
    \label{fig_pythia_noMPI_regions_nonprompt_pt}
\end{figure*}

The yield of \jpsi from NRQCD does not depend strongly on the \mult. It is independent of the multiplicity in the Away region, and decreases slightly with increasing multiplicity in the regions separated from the flight direction of the \jpsi. As a function of the multiplicity in the flight direction of the \jpsi, first an increase of the yield with multiplicity can be observed, which then saturates and changes into a decrease.
The yield of non-prompt \jpsi increases strongly as a function of the multiplicity in the Away region, and of the multiplicity in the Toward region at mid-rapidity. It is weakly dependent on the multiplicity in the other regions.
For \jpsi from cluster collapse the increase with multiplicity is qualitatively similar to the one of non-prompt \jpsi, but slightly weaker.
These observations are in line with the expectations from the described effects: an auto-correlation of the multiplicity in the region around the \jpsi from the additional decay daughters, and an auto-correlation from the recoil jet, especially for non-prompt \jpsi, which is spread out in rapidity.

Since the auto-correlation effects are strongest for non-prompt \jpsi, the multiplicity dependence is further investigated in transverse momentum intervals (\fig{fig_pythia_noMPI_regions_nonprompt_pt}). The dependence on the multiplicity in regions where no auto-correlation effects are expected, the Transverse region, and the Toward region at forward rapidity, is largely \pt independent. On the other hand, in the region affected by auto-correlations, the increase with multiplicity is strongly \pt dependent, i.e. the increase is much stronger for high \pt. This is again in line with expectations, since higher-\pt beauty quarks should fragment into more particles.

\begin{figure*}[th!]
  \centering
    \includegraphics[width=0.75\linewidth]{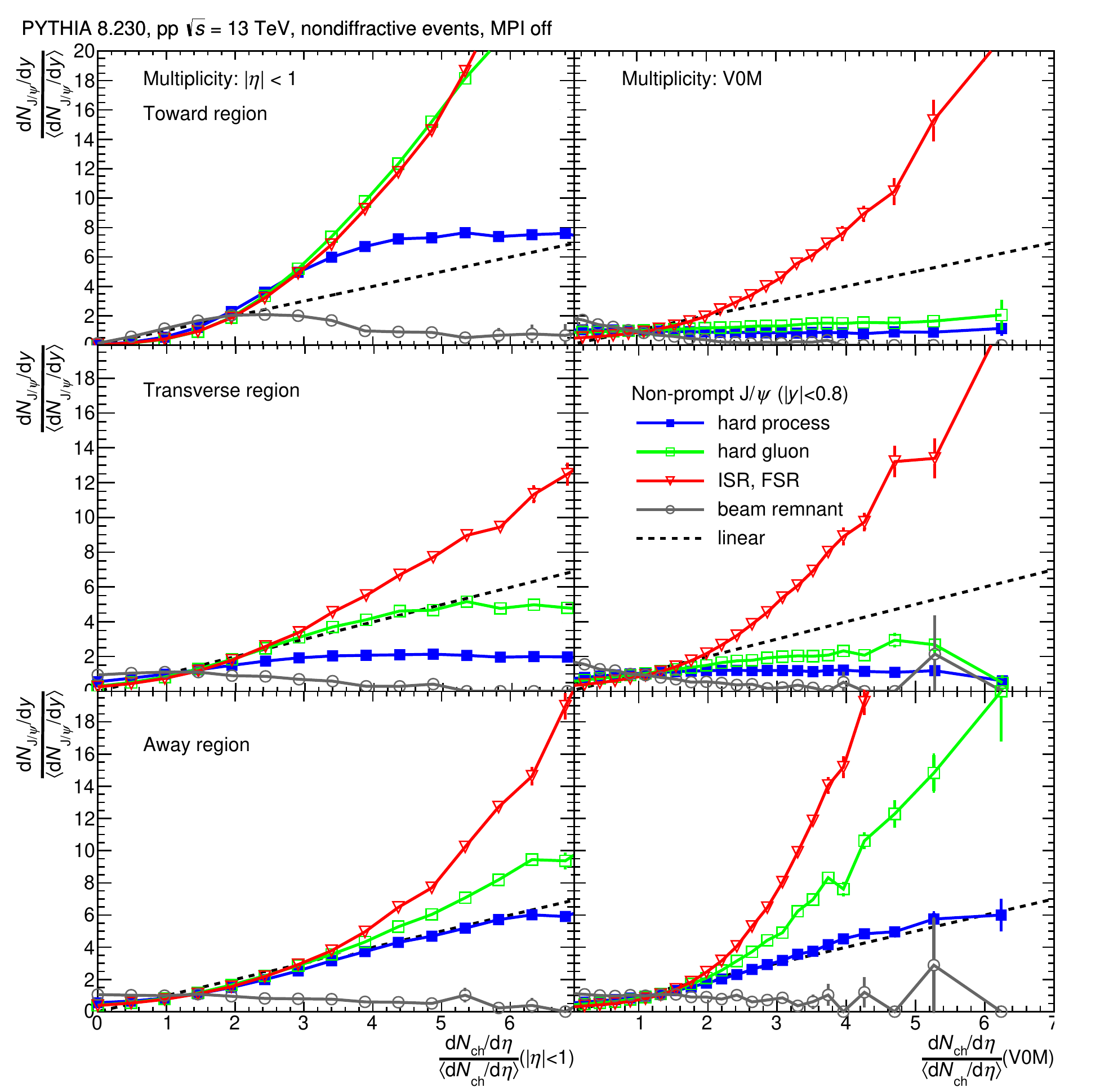} 
  \caption{Multiplicity dependence of non-prompt \jpsi produced at mid-rapidity in PYTHIA8.230 with \mpi switched off, for different production mechanisms of the beauty quark. 
    The panels are as in Fig.~\ref{fig_pythia_noMPI_regions_sources}.
  }
    \label{fig_pythia_noMPI_regions_nonprompt_processes}
\end{figure*}

Based on the auto-correlation arguments, in principle non-prompt \jpsi production should be independent of the multiplicity in the Transverse region, and of the multiplicity in the Toward region at different rapidity. However, also in these cases a slight increase with multiplicity is observed. This fact can be further investigated by splitting the non-prompt \jpsi yield depending on the production process of the initial beauty quark. As discussed earlier, this can be a hard interaction, a gluon splitting --- either of a gluon produced in a hard interaction or in ISR or FSR --- or it can come from the beam remnant. The multiplicity dependence for these different cases is shown in \fig{fig_pythia_noMPI_regions_nonprompt_processes}. If the beauty quark is produced in a hard interaction, the multiplicity dependence is as expected from auto-correlation effects only. The increase with multiplicity is present in the Toward region at the same rapidity of the \jpsi and in the Away region. In the latter case, the increase is also observed as a function of the multiplicity at forward rapidity. In these hard processes, the beauty quark and antiquark are produced back-to-back so the influence of the recoil jet is very clear, leading to the discussed auto-correlation pattern. 

For beauty quark-antiquark pairs from gluon splitting, where the gluon is produced in a hard process, the multiplicity dependence is also spread out into the other regions. In this case, the \jpsi yield also increases as a function of the multiplicity in the Transverse region. This is not unexpected, since the topology of the produced partons is different from a back-to-back leading order topology. The beauty quark is produced together with a beauty antiquark in the gluon splitting process with a small opening angle. The non-zero opening angle results in the slight dependence on the multiplicity in the Transverse region. Additionally, the gluon is produced back-to-back with another parton in the hard scattering process, producing the recoil jet signature.
In the case of the splitting of a gluon from  ISR or FSR, the increase with multiplicity is spread over all regions. This is expected, since the ISR/FSR gluon can have a large opening angle to the partons produced in the hard interaction which should produce the bulk of the multiplicity in the event.
For non-prompt \jpsi from a beauty quark of the beam remnant, the multiplicity dependence is much weaker. The reason is the lower transverse momentum of this contribution.

\begin{figure*}[th!]
  \centering
    \includegraphics[scale=0.75]{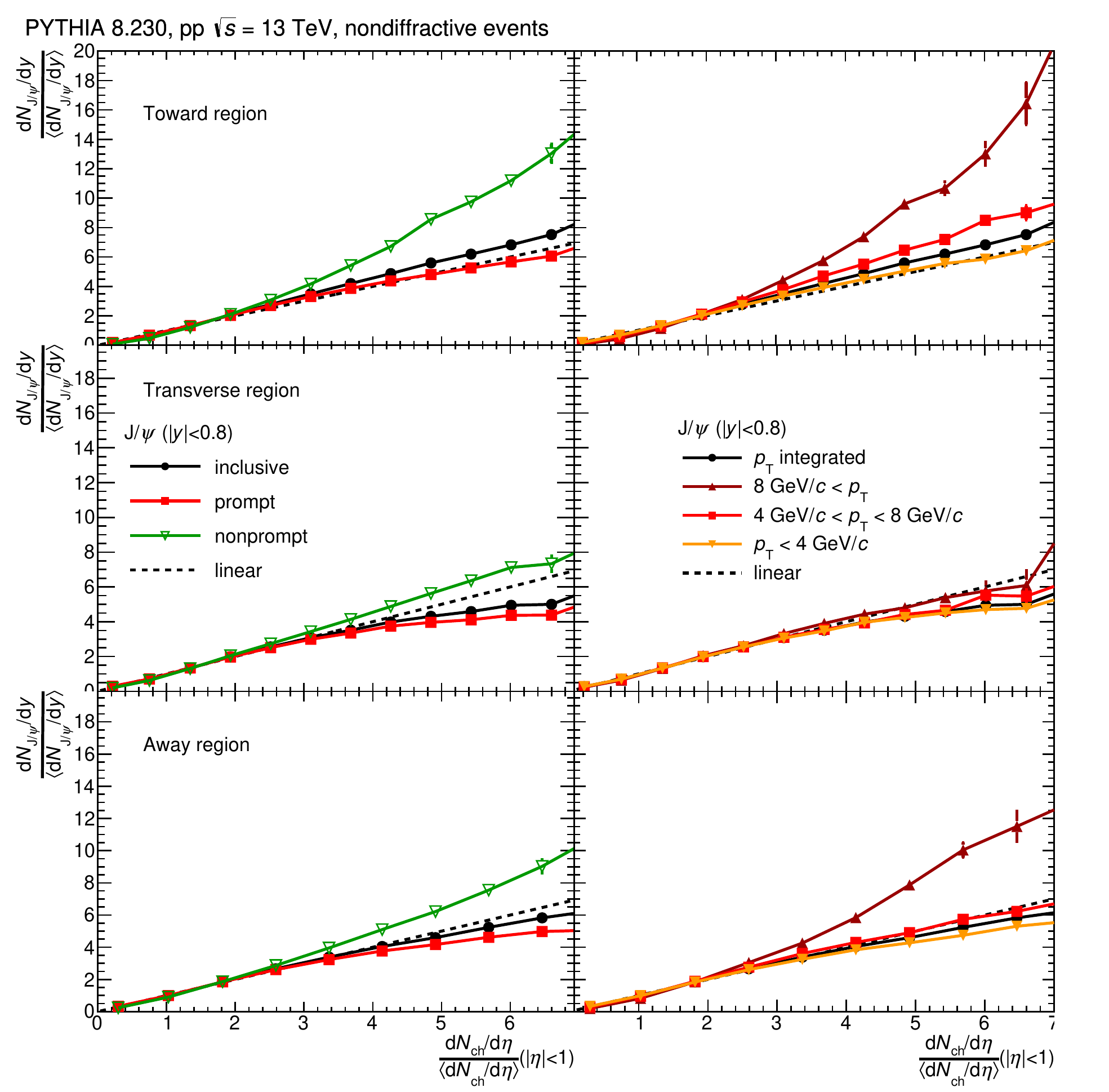} 
  \caption{Self-normalized \jpsi yield at mid-rapidity as a function of self-normalized charged-particle multiplicity at mid-rapidity in PYTHIA8.230. 
    The panels are as in Fig.~\ref{fig_pythia_noMPI_regions_sources}.
  }
    \label{fig_pythia_transverse_full}
\end{figure*}

From these observations in events without MPI it is concluded that the multiplicity dependence of \jpsi production is affected by auto-correlation effects.
In the full simulation, i.e. with inclusion of the MPI mechanism, the observed auto-correlation patterns are preserved, as shown in \fig{fig_pythia_transverse_full}. 
However, in the Transverse region the  prompt \jpsi yield follows
approximately the MPI-CR-baseline and it grows slightly stronger than linear for non-prompt \jpsi production.

\noindent A stronger-than-linear increase for non-prompt \jpsi is observed when measured as a function of the multiplicity in the Toward region (top panel), or the multiplicity in the Away region (bottom panel). In these cases also the stronger increase for higher \pt inclusive \jpsi is observed.

An interesting recent experimental observation is the production of prompt and non-prompt \jpsi inside jets at forward rapidity \cite{lhcb_jpsi_jets} and mid-rapidity \cite{jpsi_jets_cms} in pp collisions by the LHCb and CMS collaborations, respectively. It was found that the momentum fraction carried by prompt \jpsi inside jets is significantly lower than what is predicted from PYTHIA8. In other words prompt \jpsi are observed to be much less isolated than predicted. As a consequence, auto-correlation effects for prompt \jpsi are likely underestimated in PYTHIA8, which might explain the observed disagreement with the self-normalized yield of inclusive \jpsi as a function of multiplicity \cite{preliminary}.

\begin{figure*}[ht!]
\begin{minipage}[b]{0.5\linewidth}
\centering
  \includegraphics[width=0.9\linewidth]{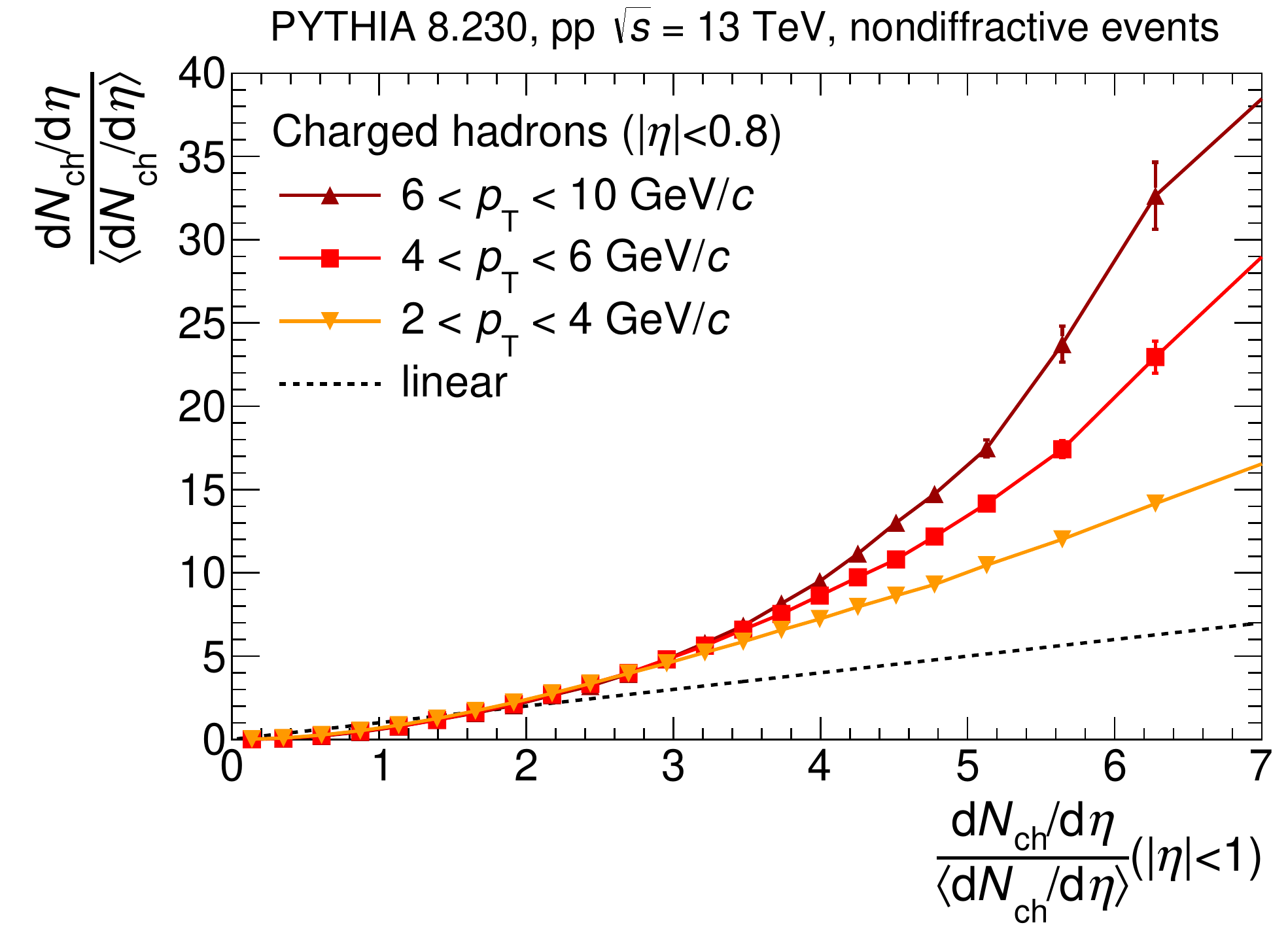} \end{minipage}
\hspace{0.1cm}
\begin{minipage}[b]{0.5\linewidth}
\centering
     \includegraphics[width=0.9\linewidth]{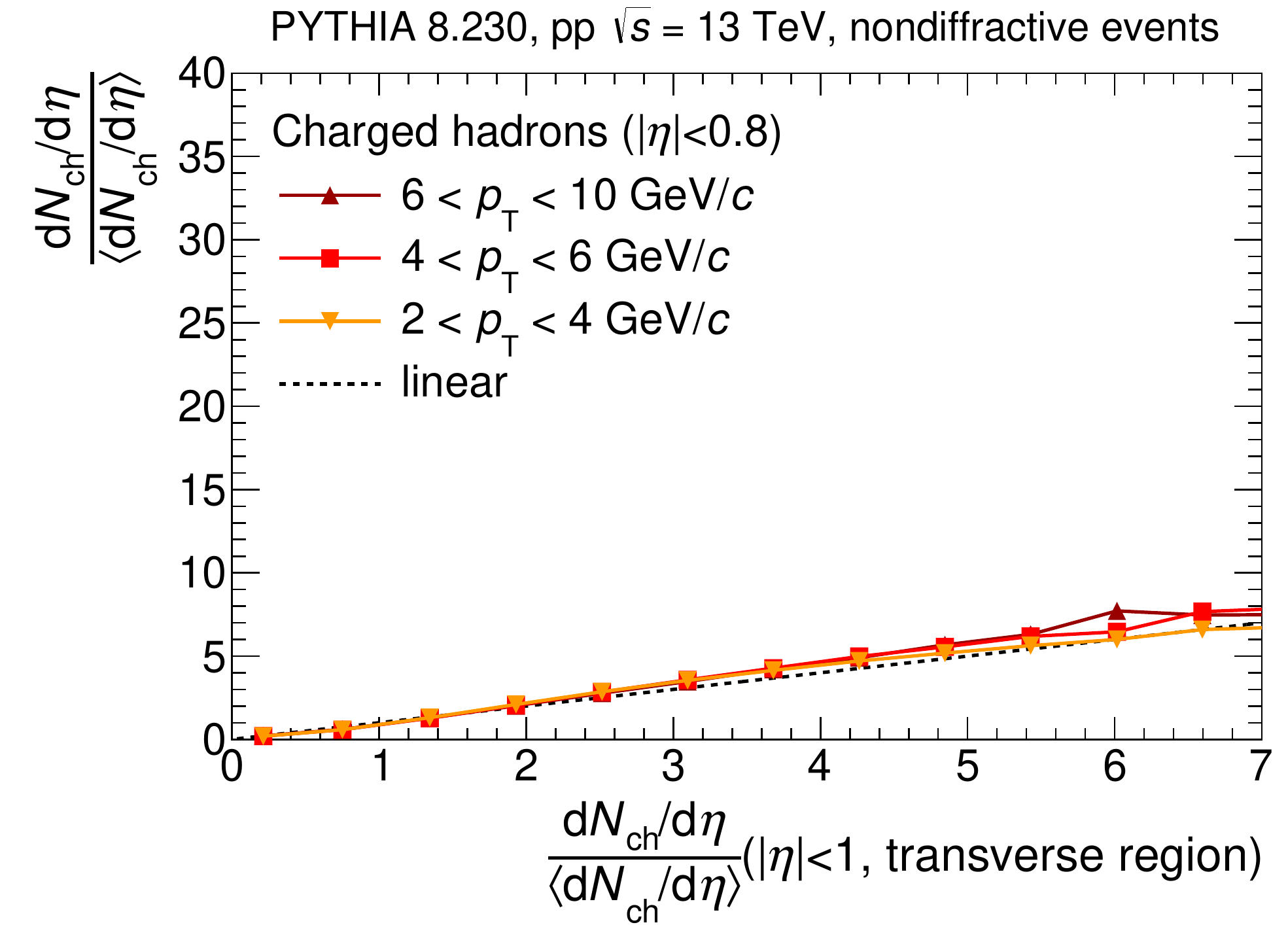} 
\end{minipage}%
\caption{Mid-rapidity hadron production as a function of \mult at mid rapidity in different \pt intervals (left) and as a function of the charge-particle multiplicity in the Transverse region (right). }
    \label{fig_pythia_hadrons}
\end{figure*}

We also studied the self-normalized yield of high-\pt charged hadrons as a function of the charged-particle multiplicity at mid-rapidity. These results are shown in the left panel of Fig.~\ref{fig_pythia_hadrons} for different \pt intervals. They reflect the behaviour observed in experimental measurements and the one previously discussed for inclusive \jpsi production. A stronger than linear increase is observed and the non-linearity
rises with the \pt of the hadron. 
Moreover, the self-normalized hadron yields have been computed as a function of the charged-particle multiplicity measured in the Transverse region as shown in the right panel of Fig. \ref{fig_pythia_hadrons}.
As observed for the \jpsi, also for the hadrons the increase with multiplicity becomes closer to linear and the \pt dependence is fully removed.
However, the deviations from the MPI-CR-baseline show that auto-correlations have still an influence.
These studies confirm our interpretation of the results for the \jpsi, and hence our conclusion that the stronger-than-linear increase (consequently also stronger than the MPI-CR-baseline), and its \pt dependence, is fully driven by auto-correlation effects.

\section{Conclusion}

From the studies reported in this paper, it is observed that PYTHIA8 qualitatively reproduces the experimentally observed stronger-than-linear increase of \jpsi production as a function of \mult. 
From the scenario of independent MPI one expects that multiplicity increases linearly with the 
number of PI. In absence of auto-correlation effects, the increase of the yield of \jpsi\ -- and in general for all hard processes -- is weaker than linear for multiplicities exceeding about three times the mean multiplicity. The trend in this region is caused by the interplay between the multiplicity fluctuations of individual PI and the steeply falling MPI probability distribution.
Moreover, we show that colour reconnection, as implemented in PYTHIA8, cannot be responsible for a stronger than 
linear increase.
A small fraction of \jpsi are formed from charm quarks produced independently. 
This contribution scales roughly quadratically as a function of the \mult.

Due to associated soft particle production in events containing non-prompt \jpsi, the multiplicity dependence is affected by auto-correlation effects leading to a stronger-than-linear increase.
The auto-correlations are still present if the multiplicity is measured at a rapidity separated from the signal. However, it can be removed effectively by measuring the multiplicity in the azimuthal region transverse to the \jpsi direction. The increase with multiplicity is then approximately linear, albeit still above the MPI-CR-baseline, and independent of transverse momentum.
Any additional effects such as string
overlapping or CGC effects \cite{percolation1,kopeliovich,cgc_jpsi} are not implemented in PYTHIA, hence theoretical
predictions have to take this into account by
assuming that the baseline behaviour, without these effects, is not a linear increase, but already stronger than linear. 

It has to be noted that the findings presented in this paper for non-prompt \jpsi production are equally valid for open heavy-flavour mesons in general, and the finding for prompt \jpsi can be applied to bottomonium production.

An experimental measurement of \jpsi production as a function of multiplicity in the Transverse region is thus highly advisable in order to disentangle between auto-correlation effects, which should vanish in this case, and true correlation effects between the hard probe and the underlying event as predicted several theoretical model calculations.

\section*{Acknowledgement}
The authors thank Sarah Porteboeuf-Houssais for the helpful discussions.
This work is part of and supported by the DFG Collaborative Research Centre "SFB 1225 (ISOQUANT)". 
Computational resources have been provided by the GSI Helmholtzzentrum f{\"u}r Schwerionenforschung.
 
\bibliographystyle{unsrt}
\bibliography{literatur.bib}

\begin{thebibliography}{10}
\expandafter\ifx\csname url\endcsname\relax
  \def\url#1{\texttt{#1}}\fi
\expandafter\ifx\csname urlprefix\endcsname\relax\def\urlprefix{URL }\fi
\expandafter\ifx\csname href\endcsname\relax
  \def\href#1#2{#2} \def\path#1{#1}\fi

\bibitem{nrqcd_cgc}
Y.-Q. Ma, R.~Venugopalan, {Comprehensive Description of \jpsi Production in
  Proton-Proton Collisions at Collider Energies}, Phys. Rev. Lett. 113~(19)
  (2014) 192301.
\newblock \href {http://arxiv.org/abs/1408.4075} {\path{arXiv:1408.4075}},
  \href {http://dx.doi.org/10.1103/PhysRevLett.113.192301}
  {\path{doi:10.1103/PhysRevLett.113.192301}}.

\bibitem{Ma:2018qvc}
Y.-Q. Ma, T.~Stebel, R.~Venugopalan, {J$/\psi$ polarization in the CGC+NRQCD
  approach}\href {http://arxiv.org/abs/1809.03573} {\path{arXiv:1809.03573}}.

\bibitem{alice_jpsi_mult_7tev}
B.~Abelev, et~al., {\jpsi production as a function of charged particle
  multiplicity in pp collisions at \sqrts{7}}, Phys. Lett. B712 (2012)
  165--175.
\newblock \href {http://arxiv.org/abs/1202.2816} {\path{arXiv:1202.2816}},
  \href {http://dx.doi.org/10.1016/j.physletb.2012.04.052}
  {\path{doi:10.1016/j.physletb.2012.04.052}}.

\bibitem{alice_d_mult_7tev}
J.~Adam, et~al., {Measurement of charm and beauty production at central
  rapidity versus charged-particle multiplicity in proton-proton collisions at
  \sqrts{7}}, JHEP 09 (2015) 148.
\newblock \href {http://arxiv.org/abs/1505.00664} {\path{arXiv:1505.00664}},
  \href {http://dx.doi.org/10.1007/JHEP09(2015)148}
  {\path{doi:10.1007/JHEP09(2015)148}}.

\bibitem{cms_y_mult}
S.~Chatrchyan, et~al., {Event activity dependence of Y(nS) production in
  \sqrtsnn{5.02} pPb and \sqrts{2.76} pp collisions}, JHEP 04 (2014) 103.
\newblock \href {http://arxiv.org/abs/1312.6300} {\path{arXiv:1312.6300}},
  \href {http://dx.doi.org/10.1007/JHEP04(2014)103}
  {\path{doi:10.1007/JHEP04(2014)103}}.

\bibitem{percolation1}
E.~G. Ferreiro, C.~Pajares, {High multiplicity pp events and \jpsi production
  at LHC}, Phys. Rev. C86 (2012) 034903.
\newblock \href {http://arxiv.org/abs/1203.5936} {\path{arXiv:1203.5936}},
  \href {http://dx.doi.org/10.1103/PhysRevC.86.034903}
  {\path{doi:10.1103/PhysRevC.86.034903}}.

\bibitem{kopeliovich}
B.~Z. Kopeliovich, H.~J. Pirner, I.~Potashnikova, K., K.~Reygers, I.~Schmidt,
  {\jpsi in high-multiplicity pp collisions: Lessons from pA collisions}, Phys.
  Rev. D88~(11) (2013) 116002.

\bibitem{cgc_jpsi}
Y.-Q. Ma, P.~Tribedy, R.~Venugopalan, K.~Watanabe, {Event engineering studies
  for heavy flavor production and hadronization in high multiplicity
  hadron-hadron and hadron-nucleus collisions}, Phys. Rev. D98~(7) (2018)
  074025.
\newblock \href {http://arxiv.org/abs/1803.11093} {\path{arXiv:1803.11093}},
  \href {http://dx.doi.org/10.1103/PhysRevD.98.074025}
  {\path{doi:10.1103/PhysRevD.98.074025}}.

\bibitem{epos3}
K.~Werner, B.~Guiot, I.~Karpenko, T.~Pierog, {Analysing radial flow features in
  p-Pb and p-p collisions at several TeV by studying identified particle
  production in EPOS3}, Phys. Rev. C89~(6) (2014) 064903.
\newblock \href {http://arxiv.org/abs/1312.1233} {\path{arXiv:1312.1233}},
  \href {http://dx.doi.org/10.1103/PhysRevC.89.064903}
  {\path{doi:10.1103/PhysRevC.89.064903}}.

\bibitem{pythia64}
T.~Sj\"{o}strand, S.~Mrenna, P.~Z. Skands, {PYTHIA 6.4 Physics and Manual},
  JHEP 05 (2006) 026.
\newblock \href {http://arxiv.org/abs/hep-ph/0603175}
  {\path{arXiv:hep-ph/0603175}}, \href
  {http://dx.doi.org/10.1088/1126-6708/2006/05/026}
  {\path{doi:10.1088/1126-6708/2006/05/026}}.

\bibitem{pythia8}
T.~Sj\"{o}strand, S.~Mrenna, P.~Z. Skands, {A Brief Introduction to PYTHIA
  8.1}, Comput. Phys. Commun. 178 (2008) 852--867.
\newblock \href {http://arxiv.org/abs/0710.3820} {\path{arXiv:0710.3820}},
  \href {http://dx.doi.org/10.1016/j.cpc.2008.01.036}
  {\path{doi:10.1016/j.cpc.2008.01.036}}.

\bibitem{mpi}
T.~Sj\"{o}strand, M.~van Zijl, {A Multiple Interaction Model for the Event
  Structure in Hadron Collisions}, Phys. Rev. D36 (1987) 2019.
\newblock \href {http://dx.doi.org/10.1103/PhysRevD.36.2019}
  {\path{doi:10.1103/PhysRevD.36.2019}}.

\bibitem{beam_remnants}
T.~Sj\"{o}strand, P.~Z. Skands, {Multiple interactions and the structure of
  beam remnants}, JHEP 03 (2004) 053.
\newblock \href {http://arxiv.org/abs/hep-ph/0402078}
  {\path{arXiv:hep-ph/0402078}}, \href
  {http://dx.doi.org/10.1088/1126-6708/2004/03/053}
  {\path{doi:10.1088/1126-6708/2004/03/053}}.

\bibitem{lund_string}
B.~Andersson, G.~Gustafson, G.~Ingelman, T.~Sj\"{o}strand, {Parton
  Fragmentation and String Dynamics}, Phys. Rept. 97 (1983) 31--145.
\newblock \href {http://dx.doi.org/10.1016/0370-1573(83)90080-7}
  {\path{doi:10.1016/0370-1573(83)90080-7}}.

\bibitem{color_reconnection}
S.~Argyropoulos, T.~Sj\"{o}strand, {Effects of color reconnection on $t\bar{t}$
  final states at the LHC}, JHEP 11 (2014) 043.
\newblock \href {http://arxiv.org/abs/1407.6653} {\path{arXiv:1407.6653}},
  \href {http://dx.doi.org/10.1007/JHEP11(2014)043}
  {\path{doi:10.1007/JHEP11(2014)043}}.

\bibitem{pythia82}
T.~Sj\"{o}strand, S.~Ask, J.~R. Christiansen, R.~Corke, N.~Desai, P.~Ilten,
  S.~Mrenna, S.~Prestel, C.~O. Rasmussen, P.~Z. Skands, {An Introduction to
  PYTHIA 8.2}, Comput. Phys. Commun. 191 (2015) 159--177.
\newblock \href {http://arxiv.org/abs/1410.3012} {\path{arXiv:1410.3012}},
  \href {http://dx.doi.org/10.1016/j.cpc.2015.01.024}
  {\path{doi:10.1016/j.cpc.2015.01.024}}.

\bibitem{monash}
P.~Z. Skands, S.~Carrazza, J.~Rojo, {Tuning PYTHIA 8.1: the Monash 2013 Tune},
  Eur. Phys. J. C74~(8) (2014) 3024.
\newblock \href {http://arxiv.org/abs/1404.5630} {\path{arXiv:1404.5630}},
  \href {http://dx.doi.org/10.1140/epjc/s10052-014-3024-y}
  {\path{doi:10.1140/epjc/s10052-014-3024-y}}.

\bibitem{alice_primaries}
\href{https://cds.cern.ch/record/2270008}{{The ALICE definition of primary
  particles}}.
\newline\urlprefix\url{https://cds.cern.ch/record/2270008}

\bibitem{VZERO}
E.~Abbas, et~al., {Performance of the ALICE VZERO system}, JINST 8 (2013)
  P10016.
\newblock \href {http://arxiv.org/abs/1306.3130} {\path{arXiv:1306.3130}},
  \href {http://dx.doi.org/10.1088/1748-0221/8/10/P10016}
  {\path{doi:10.1088/1748-0221/8/10/P10016}}.

\bibitem{atlas_nch_13tev}
G.~Aad, et~al., {Charged-particle distributions in \sqrts{13} pp interactions
  measured with the ATLAS detector at the LHC}, Phys. Lett. B758 (2016) 67--88.
\newblock \href {http://arxiv.org/abs/1602.01633} {\path{arXiv:1602.01633}},
  \href {http://dx.doi.org/10.1016/j.physletb.2016.04.050}
  {\path{doi:10.1016/j.physletb.2016.04.050}}.

\bibitem{pythia_mpi}
T.~Sj\"{o}strand, {The Development of MPI Modelling in PYTHIA. }\href
  {http://arxiv.org/abs/1706.02166} {\path{arXiv:1706.02166}}.

\bibitem{pythia82_onia}
T.~Sj\"{o}strand, Onia processes,
  \url{http://home.thep.lu.se/~torbjorn/pythia82html/OniaProcesses.html}.

\bibitem{charm_string}
E.~Norrbin, T.~Sj\"{o}strand, {Production mechanisms of charm hadrons in the
  string model}, Phys. Lett. B442 (1998) 407--416.
\newblock \href {http://arxiv.org/abs/hep-ph/9809266}
  {\path{arXiv:hep-ph/9809266}}, \href
  {http://dx.doi.org/10.1016/S0370-2693(98)01244-1}
  {\path{doi:10.1016/S0370-2693(98)01244-1}}.

\bibitem{alice_nonprompt}
B.~Abelev, et~al., {Measurement of prompt \jpsi and beauty hadron production
  cross sections at mid-rapidity in pp collisions at \sqrts{7}}, JHEP 11 (2012)
  065.
\newblock \href {http://arxiv.org/abs/1205.5880} {\path{arXiv:1205.5880}},
  \href {http://dx.doi.org/10.1007/JHEP11(2012)065}
  {\path{doi:10.1007/JHEP11(2012)065}}.

\bibitem{preliminary}
S.~G. Weber, {Measurement of \jpsi production as a function of event
  multiplicity in pp collisions at \sqrts{13} with ALICE}, Nucl. Phys. A967
  (2017) 333--336.
\newblock \href {http://arxiv.org/abs/1704.04735} {\path{arXiv:1704.04735}},
  \href {http://dx.doi.org/10.1016/j.nuclphysa.2017.06.054}
  {\path{doi:10.1016/j.nuclphysa.2017.06.054}}.

\bibitem{lhcb_jpsi_jets}
R.~Aaij, et~al., {Study of \jpsi Production in Jets}, Phys. Rev. Lett. 118~(19)
  (2017) 192001.
\newblock \href {http://arxiv.org/abs/1701.05116} {\path{arXiv:1701.05116}},
  \href {http://dx.doi.org/10.1103/PhysRevLett.118.192001}
  {\path{doi:10.1103/PhysRevLett.118.192001}}.

\bibitem{jpsi_jets_cms}
\href{http://cds.cern.ch/record/2318344}{{Production of prompt and nonprompt
  ${\rm J}\hspace{-.08em}/\hspace{-.14em}\psi$ mesons in jets in pp collisions
  at $\sqrt{s} = 5.02~\mathrm{TeV}$}}, Tech. Rep. CMS-PAS-HIN-18-012, CERN,
  Geneva (2018).
\newline\urlprefix\url{http://cds.cern.ch/record/2318344}

\end{thebibliography}
 
%
%
%
%

\end{document}